\documentclass{aa}
\usepackage{graphicx}
\usepackage{graphicx}
\usepackage{times}
\usepackage{epsfig}
\usepackage{longtable}
\usepackage{lscape}
\usepackage{amssymb}
\usepackage{latexsym}
\usepackage{natbib}
\usepackage{txfonts}
\def\kms{$\rm km\;s^{-1}$}

\def\kmsmpc{$\rm km\;s^{-1}\;Mpc^{-1}$}

\def\hb{H$\beta$}
\def\oiiipg{[O~{\small III}]$\,\lambda\lambda4959,5007$}
\def\oiiip{[O~{\small III}]$\,\lambda4959$}
\def\oiiig{[O~{\small III}]$\,\lambda5007$}
\def\oiii{[O~{\small III}]}
\def\nii{[N~{\small II}]}

\def\msun{M$_{\odot}$}
\begin{document}
   \title{VIMOS-VLT Integral Field Kinematics of the Giant Low Surface
  Brightness Galaxy ESO 323-G064 \thanks{Based on observations carried
  out at the European Southern Observatory (ESO 075.B-0695).}}

   \author{L. Coccato  \inst{1}
       \and R. A. Swaters \inst{2}
       \and V. C. Rubin \inst{3}
       \and S. D'Odorico \inst{4}
       \and S. S. McGaugh \inst{2}
}
   \offprints{L. Coccato, \\e-mail: lcoccato@mpe.mpg.de}

   \institute{Max-Plank-Institut f\"ur Extraterrestrische Physik, Giessenbachstra$\beta$e, D-85741 Garching bei M\"unchen, Germany\\
 \and Department of Astronomy, University of Maryland, College Park, MD 20742-2421, USA \\ 
 \and Carnegie Institution of Washington, 5241 Broad Branch Road NW, Washington, DC 20015, USA \\
 \and European Southern Observatory, Karl-Schwarzschild-Stra$\beta$e 2, 85748 Garching bei M\"unchen, Germany}

   \date{\today}

\abstract
% context heading (optional)
{}
% aims heading (mandatory)
{We have studied the bulge and the disk kinematics of the giant low
surface brightness galaxy ESO 323-G064 in order to investigate its
dynamical properties and the radial mass profile of the dark matter
(DM) halo.}
% methods heading (mandatory)
{We observed the galaxy with integral field spectroscopy (VLT/VIMOS,
in IFU configuration), measured the positions of the ionized gas by
fitting Gaussian functions to the \oiiipg\ and \hb\ emission lines,
and fit stellar templates to the galaxy spectra to determine velocity
and velocity dispersions. We modeled the stellar kinematics in the
bulge with spherical isotropic Jeans models and explored the
implications of self consistent and dark matter scenarios for NFW and
pseudo isothermal halos.}
% results heading (mandatory)
{In the bulge-dominated region, $r<5''$, the emission lines show
multi-peaked profiles. The disk dominated region of the galaxy,
$13''<r<30''$, exhibits regular rotation, with a flat rotation curve
that reaches $248 \pm 6$ \kms. From this we estimate the total
barionic mass to be $M_{bar} \sim 1.9 \cdot 10^{11}$ \msun\ and the
total DM halo mass to be $M_{DM} \sim 4.8 \cdot 10^{12}$ \msun. The stellar
velocity and velocity dispersion have been measured only in the
innermost $\approx 5''$ of the bulge, and reveal a regular rotation
with an observed amplitude of 140 \kms\ and a central dispersion of
$\sigma=180$ \kms. Our simple Jeans modeling shows that dark matter is
needed in the central $5''$ to explain the kinematics of the bulge,
for which we estimate a mass of $\approx (7 \pm 3) \cdot 10^{10}$
\msun. However, we are not able to disentangle different DM
scenarios. 
The computed central mass density of the bulge of ESO
323-G064 resembles the central mass density of some high
surface brightness galaxies, rather than that of low surface brightness
galaxies.}
% conclusions heading (optional), leave it empty if necessary 
{}

\titlerunning{2D Kinematics of the Giant LSB galaxy ESO 323-G064}

\authorrunning{Coccato et al.}  

\maketitle

\titlerunning{}

\authorrunning{}   

\maketitle
\section{Introduction}

Galaxy properties define a continuum, in size, luminosity, mass, and
surface brightness. Low surface brightness (LSB) galaxies have
significantly extended the range in surface brightness over which
galaxies can be studied, and thus they have received a great deal of
attention in studies of dark matter (e.g. \citealt{Pfenniger+94,
deBlok+97, McGaugh+01, Swaters+00, Swaters+03, Kuzio+08}), stellar
content (e.g. \citealt{deBlok+95, Bothun+97, Boissier+08}) and gas
content (e.g. \citealt{ONeil+03, Pizzella+08b}). Because of their
unique properties, LSB galaxies also play a significant role in our
understanding of the universe. Their distribution in the field and in
galaxy clusters will help us to understand the bright and dark matter
distribution in the local neighborhood and in the universe; their
number and density distribution may be relevant to the study of damped
Ly-$\alpha$ absorption seen against quasars (e.g. \citealt{Jimenez+99,
Oneil02}).  Like high surface brightness galaxies, LSB galaxies also
span a range in properties, such as size, mass, and bulge size. For
example, \citet{Bothun+90} discovered a class of giant LSB galaxies,
and \citet{Beijersbergen+90} studied a sample of bulge-dominated LSB
galaxies. To date, very little is known of the properties of
giant-sized or bulge-dominated LSB galaxies even though they can
enhance our understanding of galaxy formation and evolution by
providing an opportunity to sample galaxies in a hitherto almost
unexplored regime of galaxy properties. Several scenarios have been
proposed to explain their formation: density peak in voids
\citep{Hoffman+92}, bar instability \citep{Noguchi01,Mayer+04}, or
secular evolution from ring galaxies \citep{Mapelli+08}.
  
To fill the gap between regular and giant LSBs, Swaters, Rubin, \& 
McGaugh (SRM in preparation) have recently completed a study of the kinematics
of a sample of bulge dominated LSB galaxies. Their aim is to study the
dark matter properties and to determine whether bulge-dominated LSB galaxies
are dark matter dominated like other LSB galaxies (\citealt{deBlok+97,
Swaters+03}).

In this paper we show additional results obtained for the giant LSB galaxy
ESO 323-G064, selected from the SRM project. ESO 323-G064 has an
heliocentric velocity of 14830 \kms\ at a distance of  194 Mpc
(assuming $H_0=75$ \kmsmpc\ and a galactocentric correction of $-278$
\kms).  On its digital sky survey and 2MASS images, ESO 323-G064 has
a bright, compact bulge, surrounded by a low surface brightness
disk. There is some evidence of a weak bar component. The long slit
spectra of ESO 323-G064 obtained (SRM, in preparation)
with the 6.5m Baade Telescope at LCO, show strong, double-peaked
emission from H$\alpha$, \nii, [SII], [OI] in the nuclear region.
The large extent (47 kpc) of the H$\alpha$ and the large peak-to-peak
rotation velocity of 445 \kms\ (on the plane of the sky) mark ESO
323-G064 as a likely giant LSB.

In order to determine the kinematics of the stars in the
bulge-dominated region, and to compare these motions with those of the gas
in both the nucleus and at larger radii, we decided to obtain two
dimensional data using an integral field unit. This will give a
better picture of the complex gaseous kinematics especially the double
peaked emission revealed by the long slit data. Optical
two-dimensional observations in LSB galaxies have been obtained in the
past (i.e. \citealt{Swaters+03,Kuzio+06, Kuzio+08, Pizzella+08b})
but, so far, never for the stellar component of a giant LSB.

The paper is organized as follow: in Section \ref{sec:obs} we present
the integral field observations and discuss the data reduction
process; in Sections 3 and 4 we present the kinematical results for the
ionized gas and stellar components respectevely; in Section 5 we
present the Jeans model to the stellar kinematics and in Section 6 we
discuss the results.

\section{Observations and data reduction}
\label{sec:obs}
The integral-field spectroscopic observations were carried out in
service mode with the Very Large Telescope (VLT) at the European
Southern Observatory (ESO) in Paranal (Chile) from June to August 2005
during dark time. The Unit Telescope 3 (Melipal) was equipped with the
Visible Multi Object Spectrograph (VIMOS) in the Integral Field Unit
(IFU) configuration.
The seeing, measured by the ESO Differential Image Meteo Monitor, was
generally below $0\farcs7$ except for observations on July 1$^{\rm
st}$ during which the seeing was around $1\farcs5$ (one 30 minute
exposure) and observations on 4$^{\rm th}$ August in which the seeing
was between 0\farcs7 and 1\farcs0 (one 30 minute  exposure).

The observations were organized into 8 exposures of 30 minutes each,
divided into two different pointings with an offset of 25 arcsec.
Each telescope pointing has a field of view of $27\times27$ arcsec$^2$
and it is  recordered on a 4 CCD mosaic. We hereafter refer to
the 4 CCDs as {\it quadrants}, and we refer to the 2 observed fields
as {\it field A} and {\it field B}.  Quadrants \# 1,2,3 and 4 cover
the NE,SE,SW and NW portions of the field of view respectively.
Field A ($4\times30$ min exposures, with a few pixel dithering) covers
the NW side of the galaxy, while Field B ($4\times30$ min exposures
with a few pixel dithering) covers the SE side of the galaxy.  The
field of view of the four VIMOS quadrants was projected onto a micro
lens array. This was coupled to optical fibers which were rearranged
on a linear set of micro lenses to produce an entrance pseudo-slit to
the spectrograph. The pseudo-slit was $0\farcs95$ wide and generated a
total of 1600 spectra covering the field of view with a spatial
resolution of $0\farcs67$ per fiber. Each quadrant was equipped with
the HR-blue resolution grism ($4120 - 6210$ \AA) and a thinned,
back-illuminated EEV44 CCD with $2048\times4096$ pixels of
$15\times15$ $\mu$m$^2$.  The spectral resolution measured on the sky
emission lines is $ \lambda/\Delta \lambda \approx 1700$;
$\sigma_{instr} \approx 75$ \kms.

Together with every exposure a set of night calibration spectra were
taken: one comparison spectrum (Neon plus Argon) for the wavelength
calibration and 3 quartz lamp exposures for the flat field correction
and fibers identification.

For each VIMOS quadrant all the spectra were traced, identified, bias
subtracted, flat field corrected, corrected for relative fiber
transmission, and wavelength calibrated using the routines of the ESO
Recipe EXecution pipeline ({\tt ESOrex}) \footnote{{\tt ESOrex} and
{\tt MIDAS} are developed and maintained by the European Southern
Observatory}.
Cosmic rays and bad pixels were identified and cleaned using standard
MIDAS$\;^1$ routines.
We checked that the wavelength rebinning was done properly by
measuring the difference between the measured and predicted wavelength
for the brightest night-sky emission lines in the observed spectral
ranges \citep{Osterbrock+96}. The resulting accuracy in the
wavelength calibration is better than 5 \kms .
The intensity of the night-sky emission lines was used to correct for
the different relative transmission of the VIMOS quadrants.
The processed spectra were organized in a data cube using the tabulated
correspondence between each fiber and its position in the field of
view.
From the single exposures of every observed field we built a single
data cube. The spectra were co-added after correcting for the position
offset. The offset was determined by comparing the position of the
intensity peaks of the two reconstructed images obtained by collapsing
the data cubes along the wavelength direction. The accuracy of the
offset is $\pm 0.5$ pixel ($\simeq 0\farcs33$).  This slight
deterioration of the spatial resolution does not affect the results.

Finally, we co-added the two available data cubes (field A and field
B) by using the intensity peaks of the flux maps as a reference for
the alignment. In this way we produced a single data cube to be
analyzed in order to derive the surface brightness and two dimensional
field kinematics.

\begin{figure*}
\hbox{
  \epsfig{file=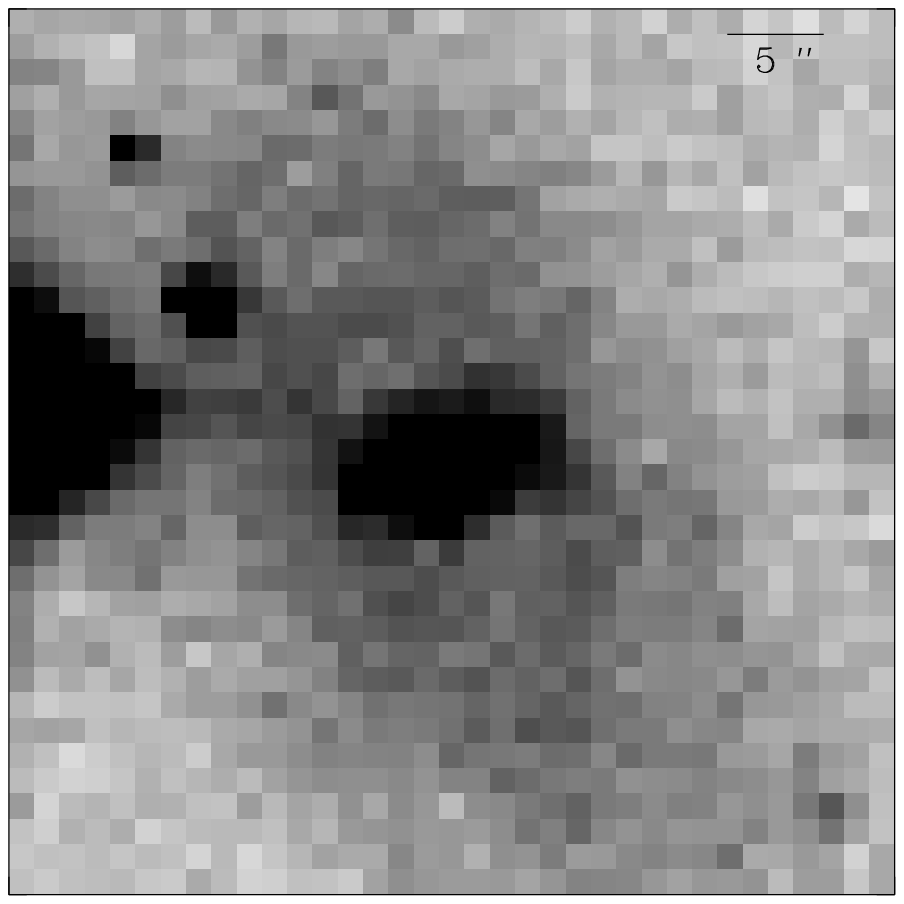,width=8.5cm,clip=}
  \epsfig{file=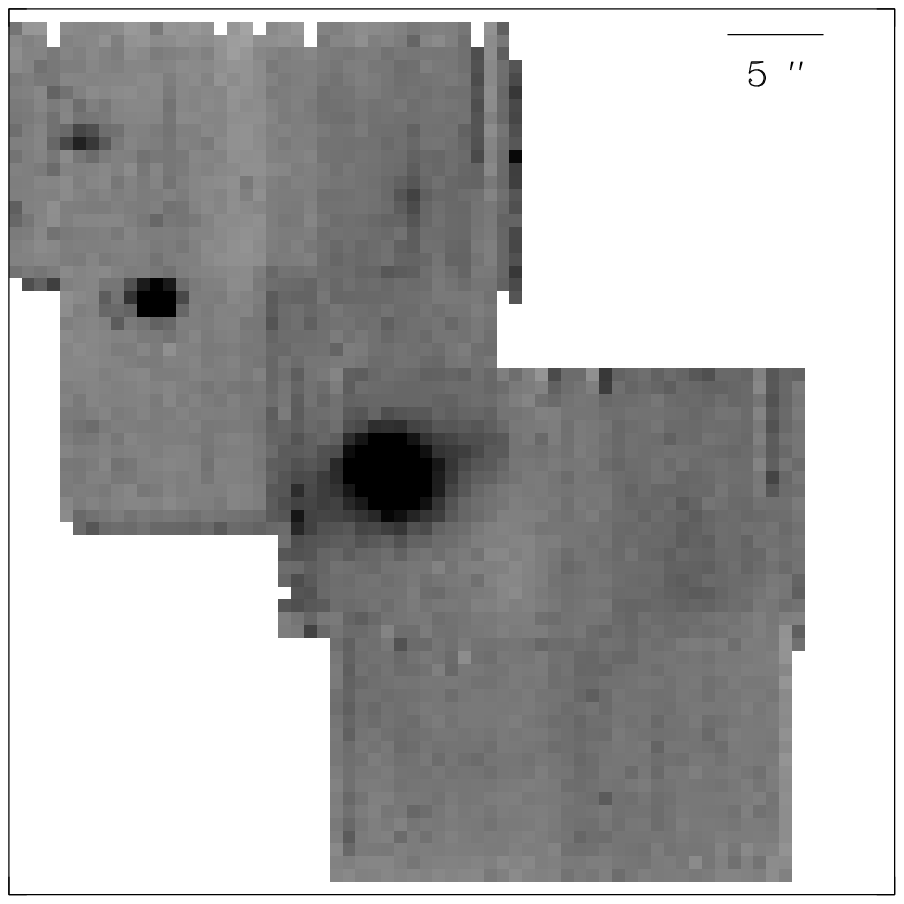,width=8.5cm,clip=}
}
\caption{{\it Left panel}: R-band image of ESO 323-G064 from the
ESO-Uppsala Galaxy surface photometry catalog
(ESO-LV, \citealt{Lauberts82}). North is top, east is left. {\it Right panel:}
reconstructed image of ESO 323-G064 from combination of 8
exposures. Scale and orientation are as in the left panel.}
%figure1a obtained by /home/coccato/VIMOS/rubin/report/richieste_referee/mkfigure2.pro cmq mkfigure
%figure1b obtained by /home/coccato/VIMOS/rubin/results/mkfigure2.pro cmq mkfigure
\label{fig:field}
\end{figure*}

In Figure \ref{fig:field} we show the two reconstructed images of both
the observed fields, obtained by collapsing the data cube along the
direction of dispersion.

\begin{figure*}
\hbox{
\psfig{file=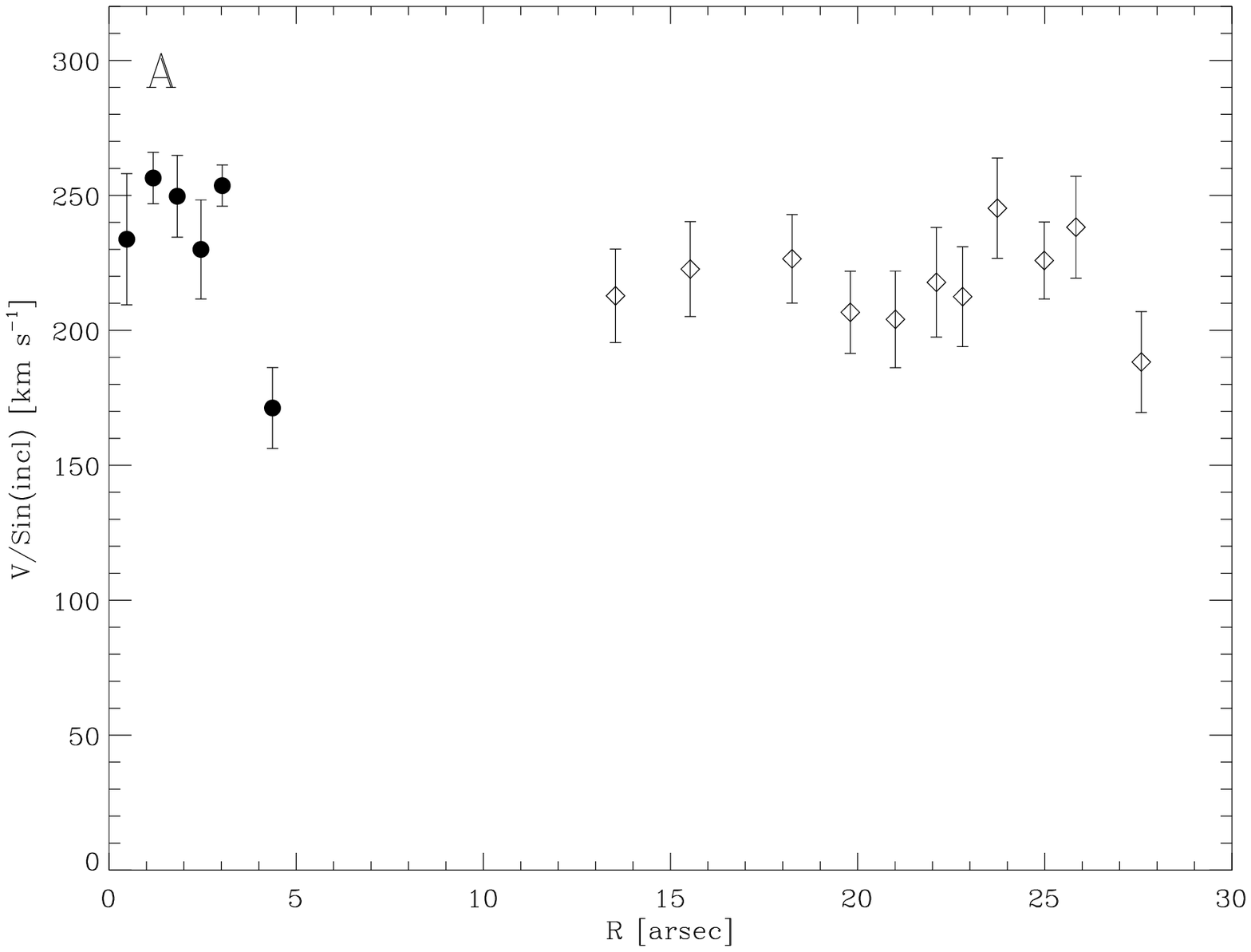,width=7.4cm,clip=} %Figure generted by ESO_JeansROT.pro, comand mkfigure. Il modello e' fittato da rubin/faint_hb/mkfig.pro comando fitta_rc (file di dati di output: myRC.dat)
\psfig{file=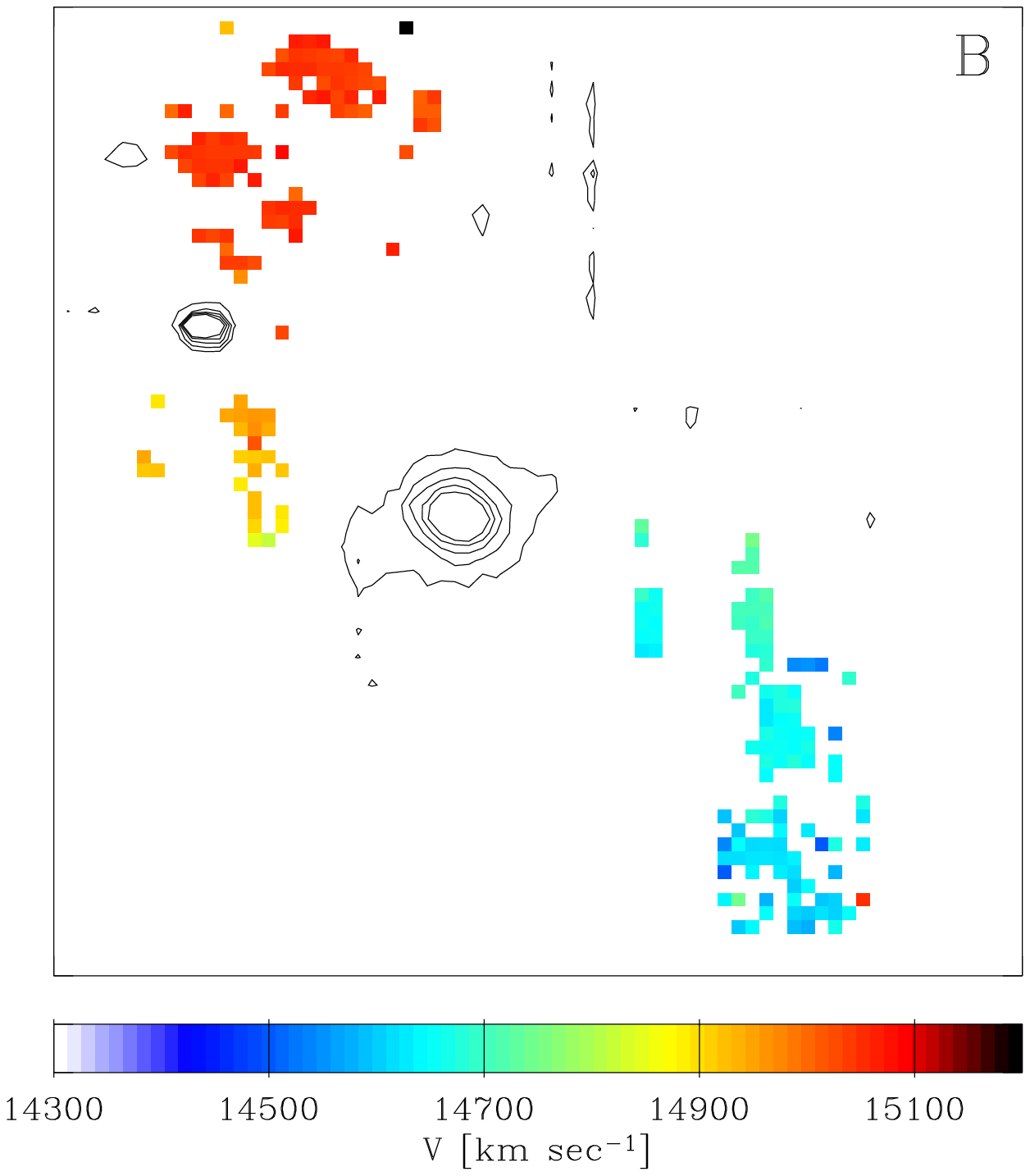,width=4.9cm,clip=}   % figura prodotta da mkfigure.pro  nella cartella report/models/vfield, comando do_figure
\psfig{file=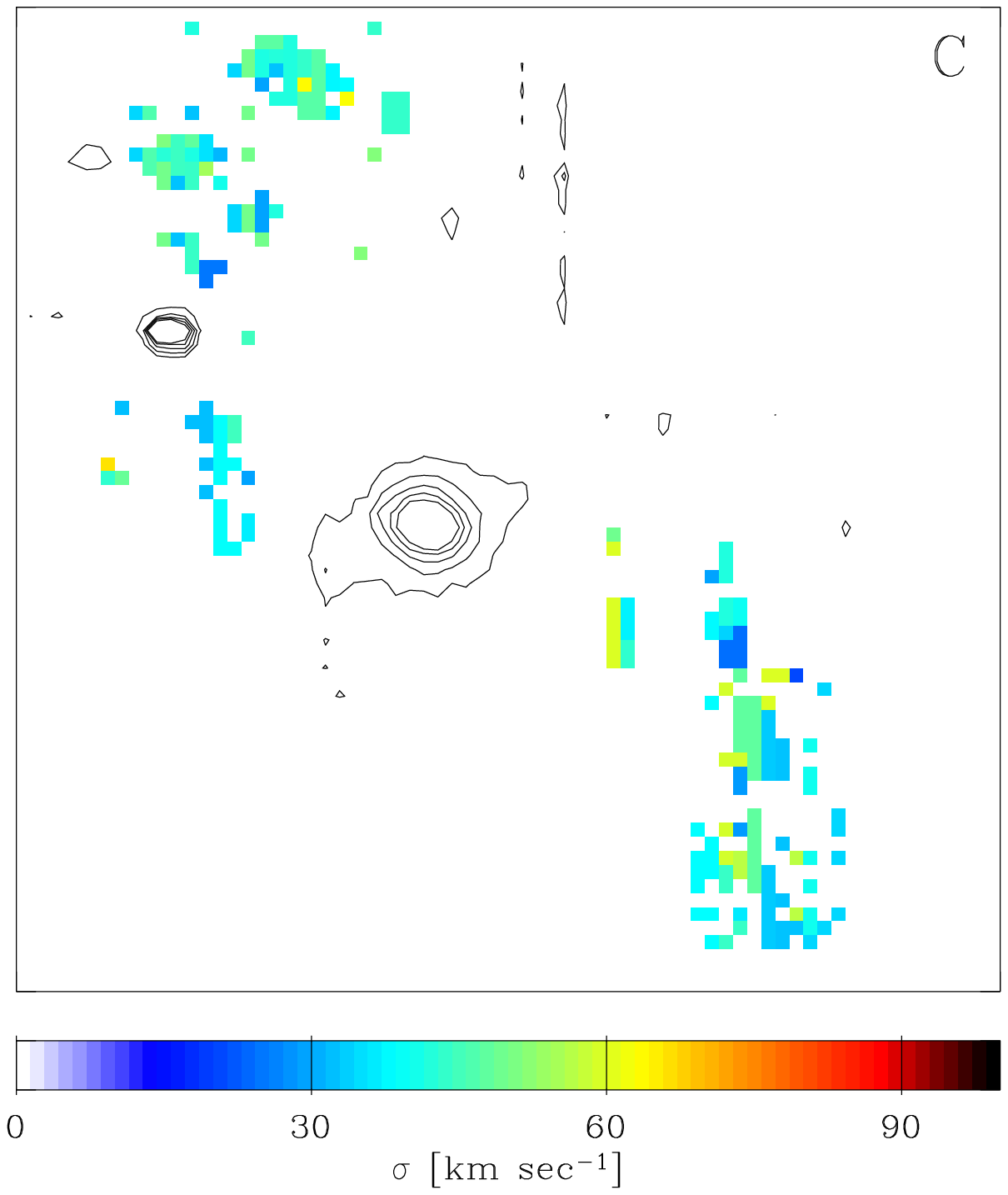,width=4.9cm,clip=}}% figura prodotta da mkfigure.pro  nella cartella report/models/vfield, comando do_figure_sigma
\caption{ {\it A:} Rotation curve of ESO 323-G064 from \hb\ emission,
    formed from the two dimensional velocity field of the disk (open
    diamonds). {\it Filled circles} are the stellar rotation circular
    velocity curve after correction for asymmetric drift (see
    \ref{sec:mass_density}).  {\it B:} Velocity field of \hb\ emission
    at large nuclear distance.  {\it C:} Velocity dispersion field of
    \hb\ emission at large nuclear distance.  In both panels {\it B}
    and {\it C}, galaxy central contours are shown, the field of view
    is $46''\times46''$, North is at top, East at left.}
\label{fig:faint_field} 
\end{figure*}   

\subsection{Sky subtraction}

The sky subtraction is a critical step in the data analysis
(especially for stellar kinematics) because 1) the galaxy absorption
lines are weak and 2) some weak sky emission lines overlap with the
galaxy emission lines.

The automatic {\tt ESOrex} pipeline evaluated and corrected the
spectra for the different efficiency of the fibers and the 4
detectors. But in order to avoid possible problems due to a poor
correction of the efficiency difference for the four quadrants, we
decided to evaluate the sky contribution for every quadrant
separately. This choice is supported by the fact that the useful
signal for the stellar kinematics is contained entirely in one
quadrant (quadrant \#3 for field A observations and quadrant \#4 for
field B observations, respectively).

For each quadrant we identified the spectra in which the galaxy's
contribution is negligible and we computed the median. Finally we
subtracted the median sky spectrum from every single spectrum of the
quadrant.

\section{Gaseous kinematics}
\label{sec.gaseous.kinematics}

ESO 323-G064 presents very bright emission lines in the nucleus
together with fainter emission lines located at larger radii.

The gas kinematics are measured by fitting to each spectrum a
background level and several Gaussian components. The number
of components is chosen depending on which and how many emission lines
are visible at the considered spectrum. No constraints are applied to
the fitting parameters: velocities, velocity dispersions and
intensities are independent.

 To determine the kinematics, we used a non-linear least-squares
minimization algorithm based on the robust Levenberg-Marquardt method
implemented by \citet{More+80}. The actual computation has been done
using the {\tt MPFIT} procedure implemented by C. B. Markwardt under
the {\tt IDL} environment\footnote{The updated version of this code is
available on http://cow.physics.wisc.edu/$\sim$craigm/idl/idl.html}.

\subsection{Ionized gas at large radii}
\label{sec:largeRadiiGas}

Faint \hb\ emission is detected at $15'' \leq r \leq 35''$. At this
distance from the center, the gas is associated with the disk
component. It reaches $\simeq 15050$ \kms\ on the NE side and $\simeq
14500$ \kms\ on the SW side. The intrinsic velocity dispersion
ranges between 30 and 50 \kms\ showing that this gaseous component is
quite cold (see Figure \ref{fig:faint_field}).

To derive a rotation curve from the two-dimensional velocity field of
the gaseous disk, we divided the gaseous disk in concentric rings,
with the same inclination ($i$) and the same orientation on the sky
($PA$). Each ring is constructed in order to contain the same numbers
(20) of data points, with velocities $v_n$ and position angles
$\phi_n$ \footnote{$\phi_n$ are measured counterclockwise starting
from the galaxy major axis.}. We also assume the gas is moving in
circular motion, therefore the velocity $V_M(R)$ in the $M-$th bin
centered at $R$ is measured by fitting the following function to the
$N=20$ points within the bin:

\begin{equation}
V(\phi,R)= V_M(R) \cdot\cos(\phi + PA) \cdot \sin(i) + V_{sys}
\label{eqn:circ_vel}
\end{equation}

The best fit results in a systemic velocity of $V_{sys}=14830 \pm 12$
\kms, a position angle on the sky of $PA=38 \pm 3$ (slightly different
from the 32 degrees derived on the ESO-LV R-band and B-band
images, an inclination of $i=61.5 \pm 0.9$ (consistent with the 60
degrees from  ESO-LV images \footnote{Images from scanned
plates can be downloaded from 
http://www.astro-wise.org/portal/aw\_datasources.shtml.}.) and a
center consistent within a pixel with the photometric center of the
bulge.  At every bin position the {\tt IDL} fitting procedure gives
the amplitude of rotation $V_M(R)$ and its error.
We show in the left panel of Figure \ref{fig:faint_field} the derived
rotation curve of ESO 323-G064. The rotation curve is relatively
constant with radius. The average rotation value (taking into account
the incliation) is $248 \pm 6$ \kms.
The result does not change if we hold the position angle
constant at $PA=32$ as derived from ESO-LV images.

Since the velocity dispersion of the ionized gas is quite small, we
can approximate the measured velocity rotation with the circular
velocity $V_C$. This allows us to estimate the total baryonic mass of
the galaxy, from the baryonic Tully Fisher relation, as described in
\citet{McGaugh05}:

\begin{equation}
M_{bar}= 50. \cdot V_C ^4. = (1.9 \pm 0.2) \cdot 10^{11} M_{\odot}
\label{eqn:total_bar_mass}
\end{equation}

In addition, we can also make a rough estimate of the total
mass $M_{DM}$ of the dark matter halo through the relation:

\begin{equation}
M_{DM}=2.52 \cdot 10^{12} \left( \frac{V_C}{200}\right)^3 =  (4.8 \pm 0.4) \cdot 10^{12} M_{\odot}
\label{eqn:total_dm_mass}
\end{equation}

To derive the formula, we followed the prescriptions by
\citet{Bryan+98} and \citet{Bullock+01} using $\Omega_\Lambda=0.7$,
$\Omega_m=0.3$, $h=0.7$, $\Delta_{c}=105.4$ and $z \approx 0.05$. The
error estimates for $M_{bar}$ and $M_{DM}$ are simply given by the
error propagation in Equations \ref{eqn:total_bar_mass} and
\ref{eqn:total_dm_mass}.

\subsection{Ionized gas in nuclear region}
 
\begin{figure}
\epsfig{file=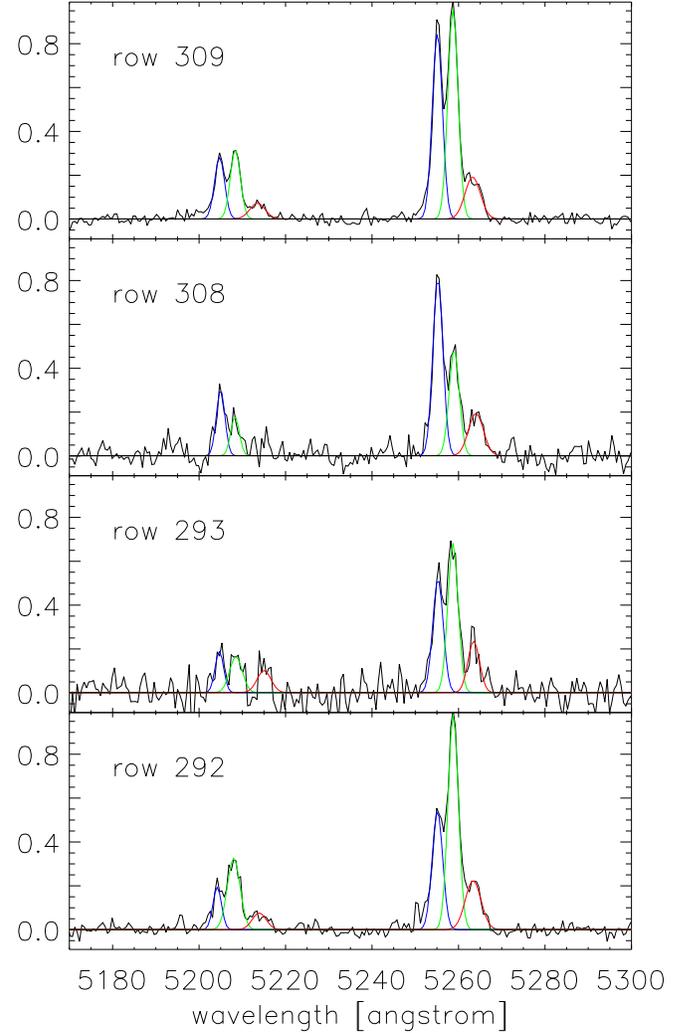,width=9cm,clip=}%,bb=50 380 540 760}
\caption{Examples of \oiii\ emission line profiles, in which the 3
peaks are visible. \hb\ emission line has been omitted in this plot
because it is very faint with respect to the \oiii. The blue, green
and red lines represent the fits to the first, second and third
emitting regions respectively.}
\label{fig:line_profile} 
\end{figure}

\begin{figure}
\vbox{
  \epsfig{file=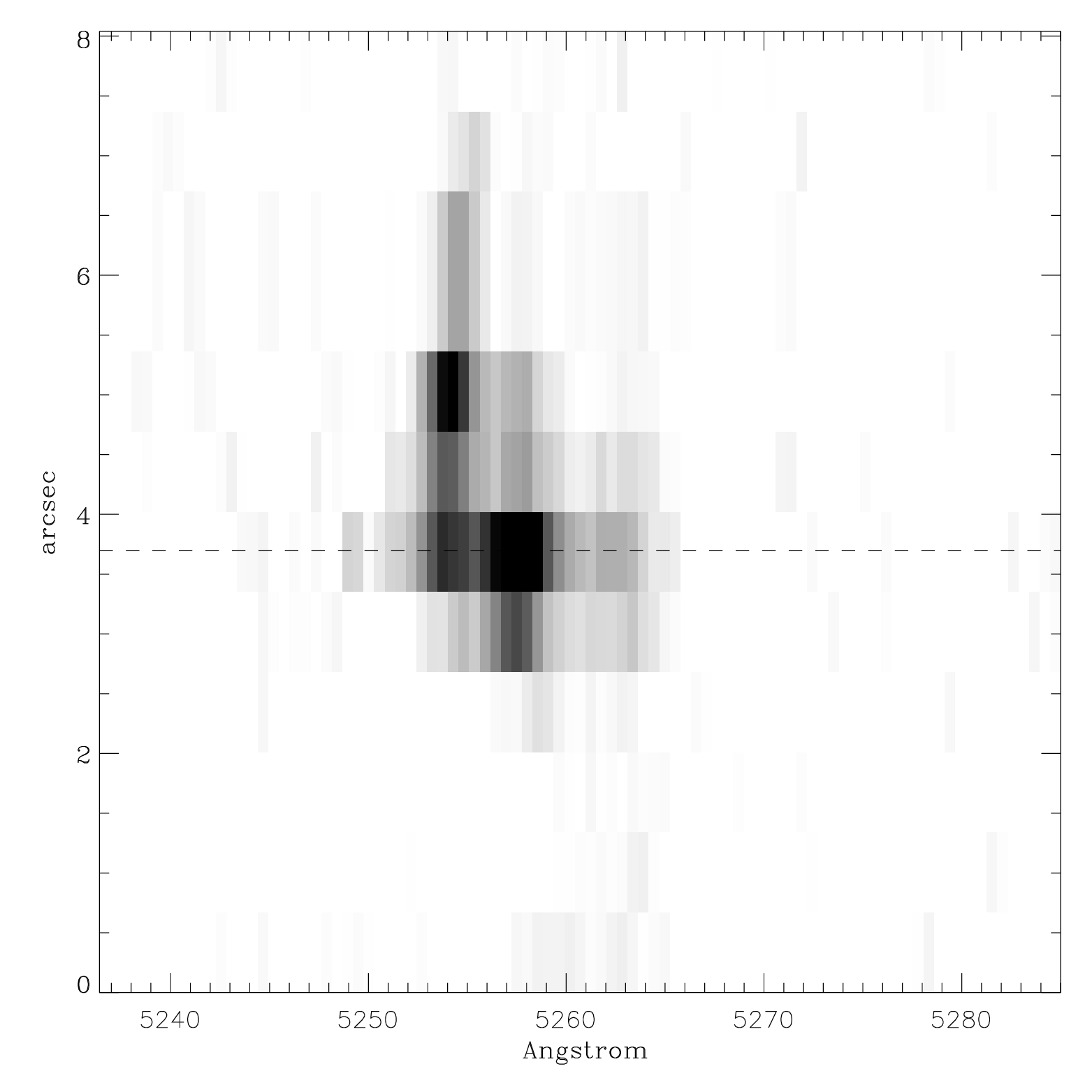,width=7.5cm,clip=}   %created by additional_figures/pvdiagram/pvdiagram.pro
  \epsfig{file=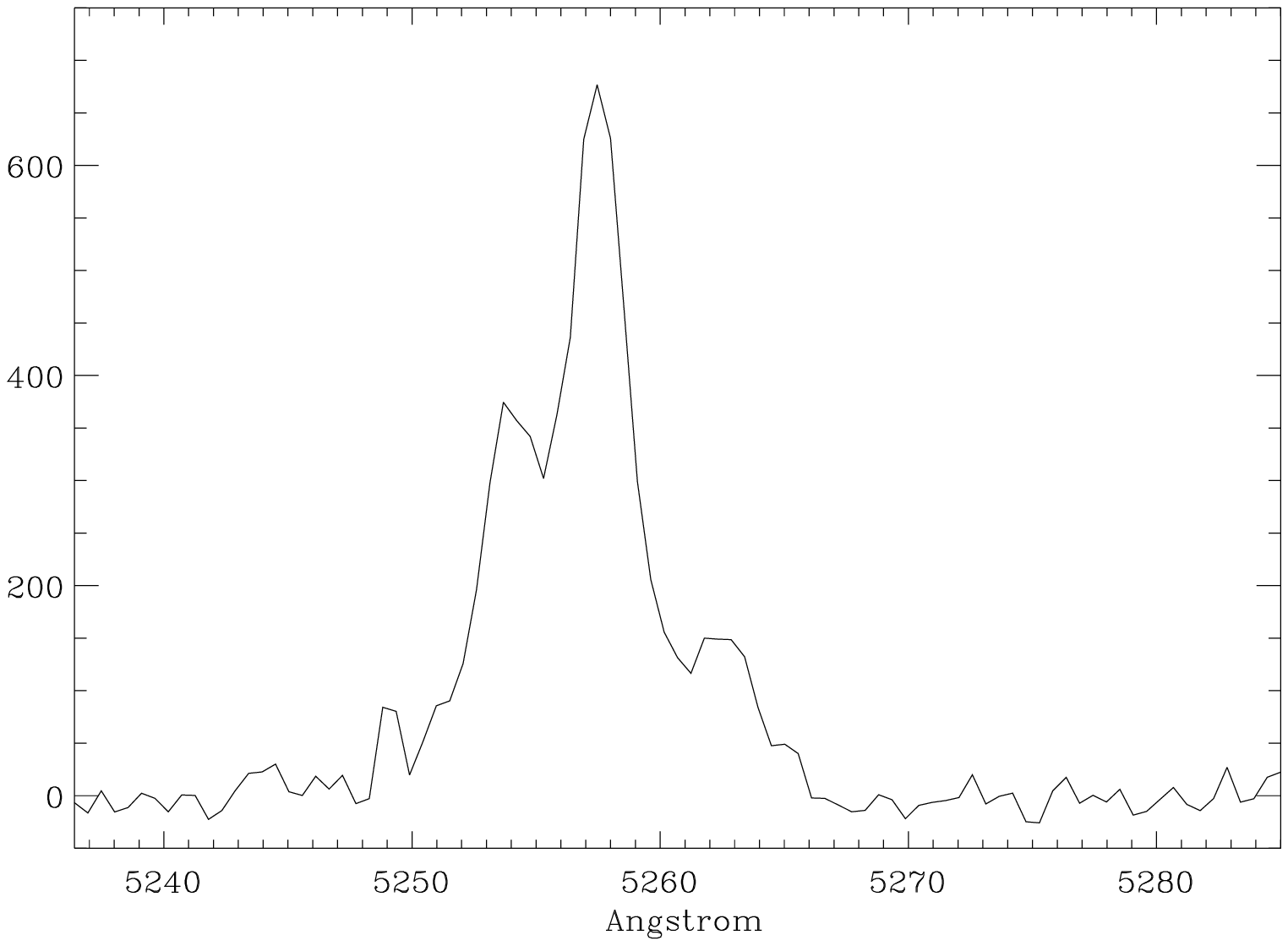,width=7.5cm,angle=0,clip=} %created by additional_figures/pvdiagram/pvdiagram.pro
}
\caption{ {\it Upper panel:} position velocity diagram towards the
galactic nucleus extracted along the major axis. {\it Lower panel:}
intensity profile of the PV diagram along the dashed line in the upper
panel. The 3 peaks of the emission line are clearly visible.}
\label{fig:line_profile2} 
\end{figure}

\hb, \oiiip\ and \oiiig\ bright emission lines are visible in the
nuclear regions within $r<5''$. Interestingly, at some locations
these emission lines are triple peaked (see Figures
\ref{fig:line_profile} and \ref{fig:line_profile2}). This complexity
is probably generated by three barely resolved emission-line
regions located close to the center but at different azimuthal
angles.

In order to have kinematical informations for all the 3 regions, we
fitted each of the the galaxy emission lines with three independent
Gaussian functions, and the results (velocity, velocity dispersion and
intensity) were compiled separately.

The velocity field of the three components do not show a regular
 rotation. Their averaged radial velocities are approximately $14539
 \pm 38$, $14726 \pm 31$, and $14992 \pm 26$ \kms. These values were
 inferred from the \oiii\ velocity field (mean value between \oiiip\
 and \oiiig ) because it is not affected by stellar absorption as is
 the \hb\ emission line. The velocity dispersion in each of the 3
 regions is between 50 and 100 \kms.

The measured FWHM of the 3 regions is about 1\farcs7, which is very
 similar to the one of the foreground star ($\approx $ 1\farcs3).  The
 maximum distance measured between the centroids of the 3 regions is
 1\farcs0.

We can exclude that the observed multi-peak line profile is an
artifact caused by the data reduction because:

\begin{enumerate}

\item The sky lines observed in each quadrant (either in the raw
spectra or in the final reduced data cube) are not multi
peaked. This eliminates instrumental effects (e.g., a shift of the
instrument during the exposure) or pipeline errors (e.g., a shift in
the dispersion direction during the data cube combining) as a source
of the triple-peaked profiles.

\item The three peaks in the galaxy emission line are present before
and after the sky subtraction and they are present also in the raw
spectra. Moreover, the few sky lines which are overlap with the galaxy
emission lines are very weak.

\item The three peaks are visible in all the galaxy emission
lines and the ratio between the intensities of the components is
(almost) the same if measured in \hb, \oiiip\ or \oiiig.

\item The peaks are visible also if we consider only the exposure with
the best seeing condition.

\end{enumerate}


\section{Stellar kinematics}
\label{sec:stellar_kinematics}

The stellar kinematics have been measured by means of the Penalized
Pixel-Fitting (ppxf) method by \citet{Cappellari+04}. We chose
a library of stellar templates from \citet{Valdes+04} provided
together with the ppfx tools.  These spectra have a spectral range of
4780 -- 5460 \AA\ and spectral resolution of 1.0 \AA\ at 5100 \AA\
($\sigma \sim 22$ \kms, which had been deteriorated in order to
match the instrumental one).

The stellar templates from this library were choosen among type K
stars, which are commonly used as kinematical templates.

Errors of the measurements ranged around $15-20$ \kms. We did several
tests using additional templates from the same library and from the
MILES library \citep{Sanchez-Blazquez+06}. Results differ from the
original values with a scatter of 10 \kms, which is consistent with
the measurement errors.

\begin{figure*}
\epsfig{file=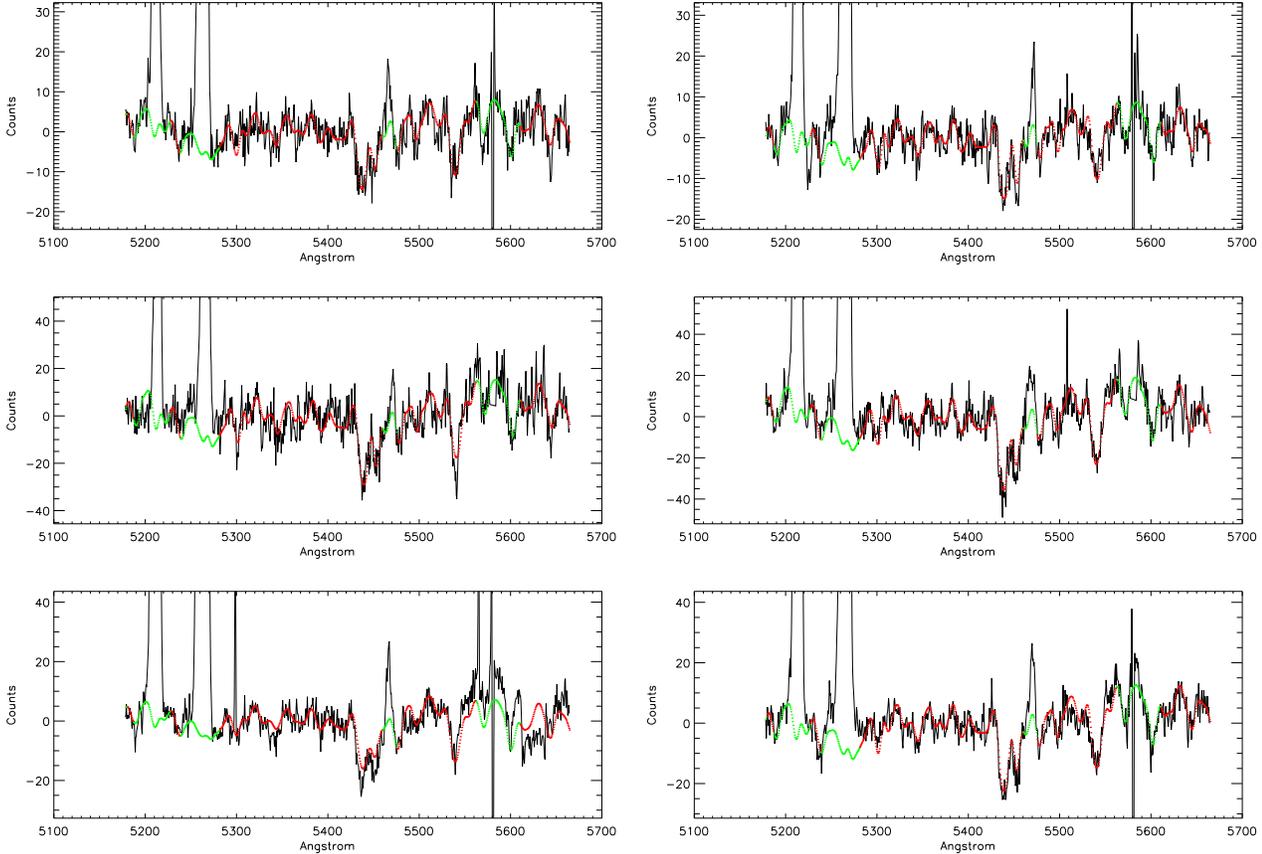,width=17cm,clip=}
\caption{Examples of the stellar kinematics fit quality. {\it Black} is
the galaxy spectrum, while {\it red} ({\it green}) represents the
portion of the stellar template which was included (excluded) from the
fit.}
\label{fig:stellar_fit_results}
\end{figure*}

No multiple peaked features in the absorption line spectra are visible
at our spectral resolution (and at our signal to noise ratio). Some
examples of the fit results are given in Figure
\ref{fig:stellar_fit_results} and the two-dimensional velocity and
velocity dispersion fields are shown in Figure \ref{fig:star}.  The
velocity of the stars ranges from $\sim 14700$ to $\sim 14900$ \kms,
and a clear pattern of rotation is visible.

%The stellar
%velocity and velocity dispersion fields are shown in Figure
%\ref{fig:star}.

The central intrinsic velocity dispersion within $1''$ is $\approx$
$180\pm10$ \kms, which, combined with the gaseous circular velocity
value $V_C= 248$ \kms\ (derived in Section \ref{sec:largeRadiiGas}),
set this galaxy close the $(V_C-\sigma_c)$ relation for LSB
galaxies \citep{Courteau+07b,Courteau+07}. According to these authors,
the predicted $V_C$ value for a bulge with $\sigma_c=180$ km/sec is
$V_C \approx 290$ \kms.

The $V/\sigma$ of the bulge is computed using the formalism introduced
by \citet{Binney05} and applied by \citet{Cappellari+07}.

\begin{eqnarray}
\left( \frac{V}{\sigma} \right)_{INTR} &=& \left( \frac{V}{\sigma}\right)_{OBS} / \sin(i) \nonumber \\ 
&=& \frac{1}{\sin(i)} \sqrt{\frac{\sum_{n=1}^N F_nV_n^2}{\sum_{n=1}^N F_n\sigma_n^2}}\approx 0.56 \pm 0.02
\end{eqnarray}
where $F_n$, $V_n$ and $\sigma_n$ are the flux (extrapolated from the
surface brightness profile), velocity and velocity dispersion measured
from the $n-th$ spectra. An inclination of 62
degrees and isotropy assumption are adopted to evaluate the
measurements in the edge-on case. Errors are calculated using classic
formulas for error propagation.

This value is consistent with the predicted value of an edge-on
isotropic oblate system \citep{Binney78} $0.4< \left(
\frac{V}{\sigma}\right)_{INTR} < 0.6$ for an intrinsic ellipticity
ranging $0.13 < \epsilon_{INTR} <0.27$. The intrinsic value for
ellipticity has been calculated from the observed one (
$0.1<\epsilon_{OBS}<0.2$, Section \ref{sec:photometry}) after
correction for inclination using the formula (see
\citealt{Binney+87}):

\begin{equation}
\epsilon_{INTR} = 1-\sqrt{1+\epsilon_{OBS} (\epsilon_{OBS} - 2)/\sin(i)}
\end{equation}

\section{Stellar dynamical modeling}
\label{sec:dynamical_modeling}

In this section we present a simple dynamical model of the stellar
velocity and velocity dispersion fields\footnote{We assume
$H_0=75$ \kmsmpc\ in the modeling.}  .

We assume the total luminous plus dark matter mass distribution of
the central galaxy bulge is spherical. The low reported value
for the ellipticity (Section \ref{sec:photometry}) allows us to
make this assumption. In addition we assume isotropy for the bulge,
given the fact that the inferred $V/\sigma$ is consistent with an
isotropic rotator, as shown in Section \ref{sec:stellar_kinematics}.

\begin{figure}
\vbox{
\epsfig{file=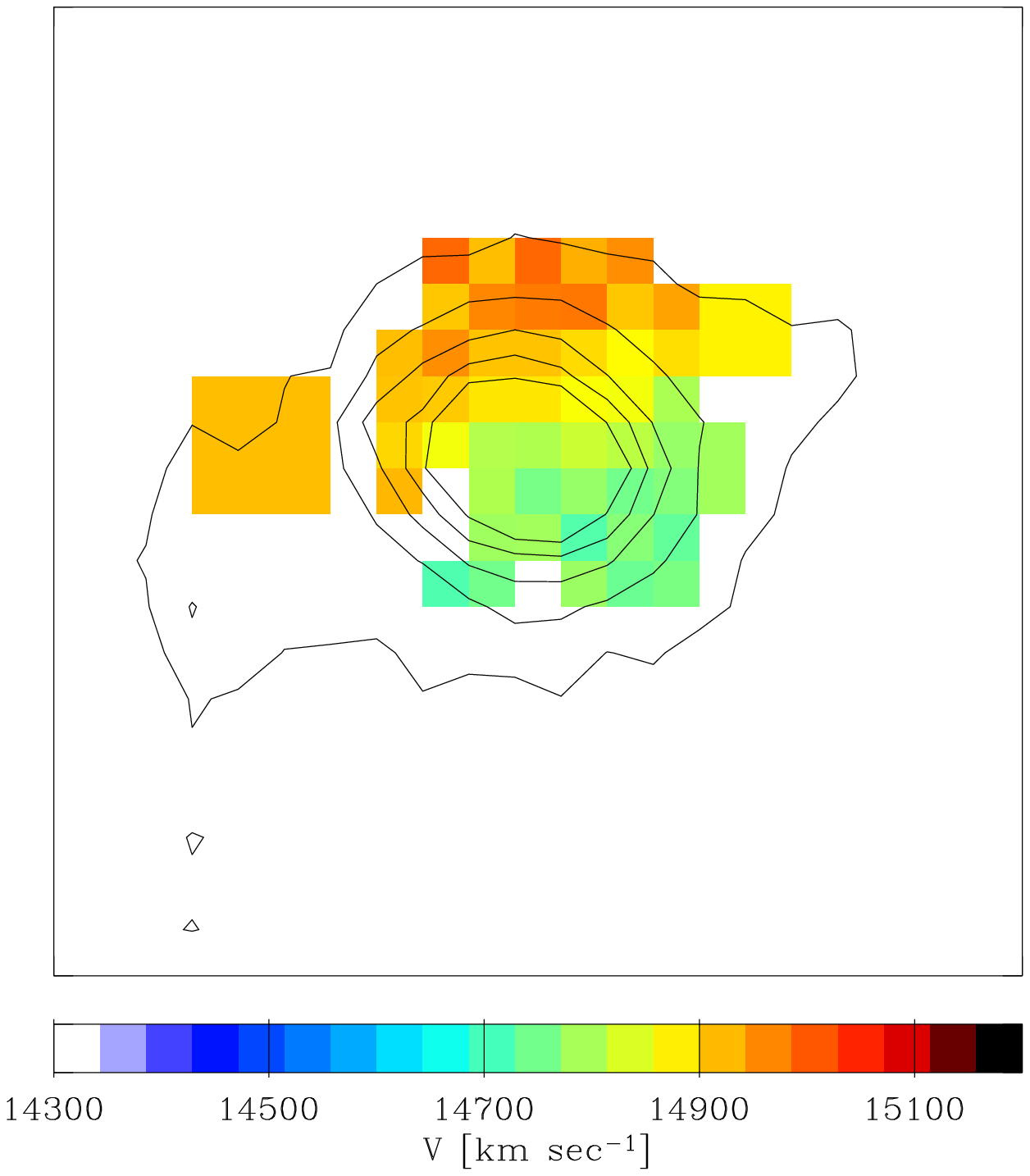,width=8cm,clip=}
 \epsfig{file=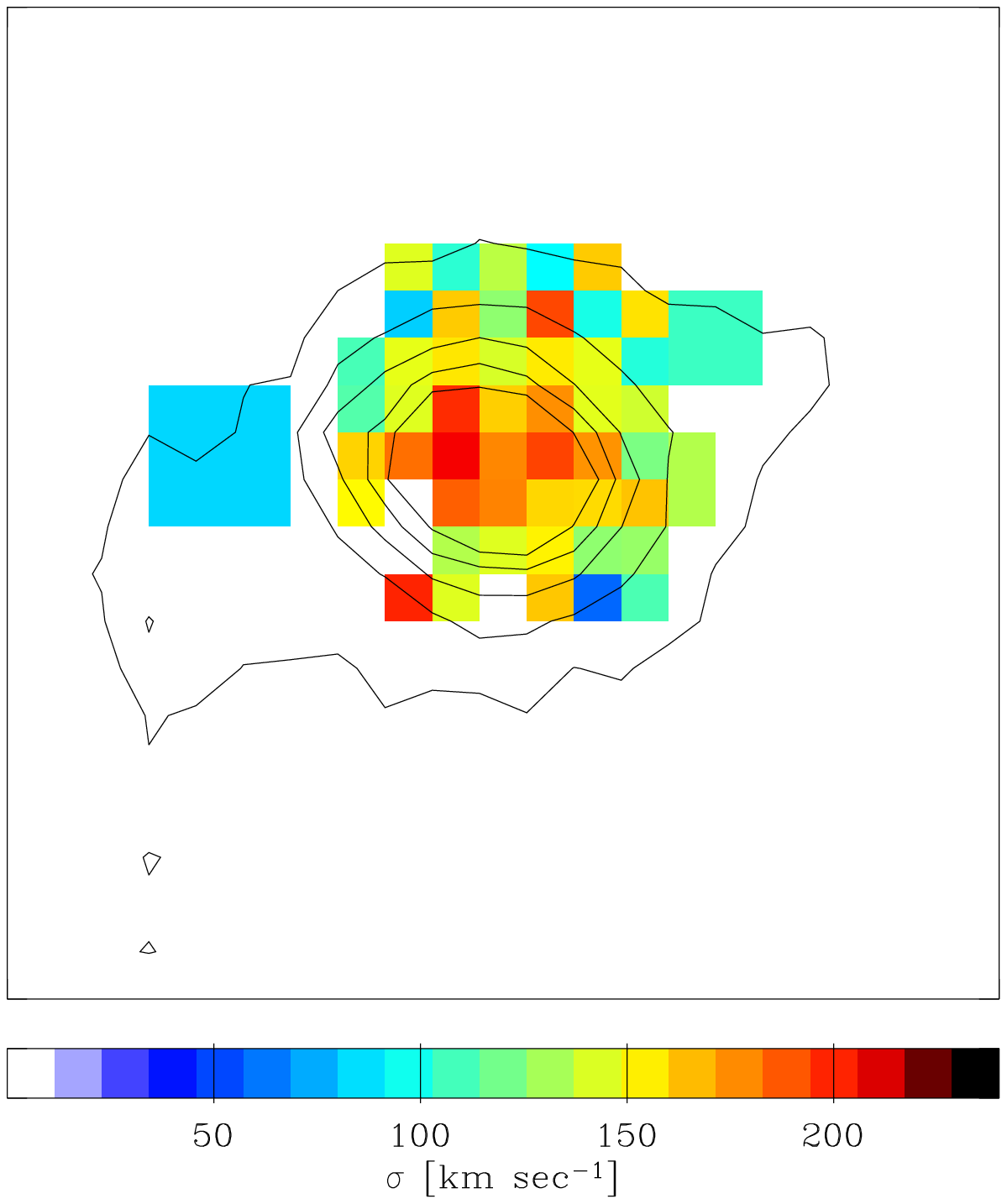,width=8cm,clip=}}
\caption{Stellar radial velocity and velocity dispersion fields, 
    with the bulge central isophotes shown. The field of view is
  $14''\times14''$, North is up, East is left.}
\label{fig:star}
\end{figure}

The main deviation from spherical symmetry in the bulge of ESO
323-G064 is the bar component, clearly visible in Figure
\ref{fig:field}. Given the limited amount of stellar kinematic data in
the bulge, it is not possible to take into account the complexity of
stellar motions in the bar potential. However, from the isophotes
shown in Figure \ref{fig:star} it is evident that the bar contribution
dominates only the outer regions of the velocity field, for $R\geq$
3\farcs5. Stellar orbits in those regions will be elongated towards
the bar direction, which is nearly orthogonal to the kinematic major
axis. This will lead to an underestimation of the stellar rotation
velocity and it will be quantified in Section
\ref{sec:model_results}.

We assume also that the disk contribution to the total mass in the
bulge region (inner $5''$) is negligible. We do not have direct
measurement of the stellar disk kinematics, but its low luminosity
(compared to the bulge) supports our assumption. Moreover, several
studies demonstrated that the disk contribution to the inner dynamics
in low surface brightness galaxies is almost insignificant
(\citealt{Swaters+00, Swaters+03}).

 The limited spatial extent of the observations and the lack of good
photometry for this galaxy do not allow us to go in more detail.

The Jeans equation in radial coordinates for such a system is (we
followed the formalism adopted by \citealt{Hui+95} and
\citealt{Peng+04}):

\begin{equation}
\frac{\rho_k(r) \sigma^2_{int}(r)}{r} = -\frac{G M(r) \rho_k(r)}{r^2} + \frac{\rho_k(r) V^2_{rot}(r)}{r}
\end{equation}
which, when solved for the intrinsic velocity dispersion $\sigma_{int}$, is:

\begin{equation}
\sigma^2_{int}(r) = \frac{1}{\rho_k(r)}  \int_r^{\infty} \rho_k(x)\frac{G M(x) -x V^2_{rot}(x)}{x^2} dx
\label{eqn.jeans}
\end{equation}
where $\rho_k$ is the mass density radial profile for the tracer of
the potential (i.e. the stars), $V_{rot} $ is the intrinsic rotation
curve of the stellar component, $M$ is the total mass of the galaxy
and $G$ is the gravitational constant.

The velocity dispersion projected on the sky is:

\begin{equation}
\sigma^2(R)=\overline{V_{LOS}^2}(R)-V_s(R)^2
\label{eqn.projected.dispersion}
\end{equation}
where $V_s$ is the projected rotation curve of the stars (see Section
\ref{sec.rotation.curve}) and  $\overline{V_{LOS}}$ is the
line-of-sight second velocity moment, projected into the sky, given
by:

\begin{equation}
\overline{V_{LOS}^2}(R)=\frac{2R}{\Sigma(R)}\int_R^{\infty}\left( \sigma^2_{int}(r)+V_{rot}^2(r)\frac{R^2}{r^2}\right)\frac{\rho_k(r)r}{\sqrt{r^2-R^2}}dr
\label{eqn.velocity.moment}
\end{equation}

\subsection{Density of the kinematical tracer}
\label{sec:photometry}

\begin{figure}
 \psfig{file=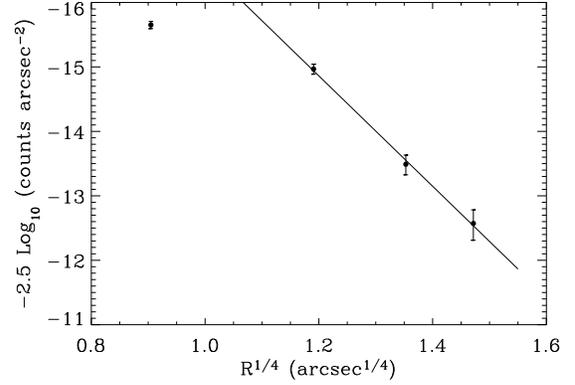,clip=,width=8cm}
\caption{{\it Dots:} Surface brightness profile obtained by collapsing
 the data cube along dispersion direction. {\it Continuous line:} de
 Vaucouleurs fit to the measured surface brightness, with $R_e=$
 0\farcs9 and $\mu_e=-16.8$ mag arcsec$^{-2}$ (arbitrary zero
 point). The innermost point was not included in the fit, since it is
 affected by the seeing.}
 \label{fig.photometry}
\end{figure}

There are no good CCD images of ESO 323-G064 available in the
literature. The better available photometry is from the ESO-LV catalog.
\footnote{See Section \ref{sec:largeRadiiGas} for the images
source.}  We performed a bulge/disk decomposition using {\tt
galfit} \citep{Peng+02}, adopting a de Vaucouleurs law for the bulge
and an exponential law for the disk. We derived an effective radius
$R_e = $ 0\farcs67 and an average ellipticity of 0.3 for the bulge,
position angle $PA=32$ and inclination $i=60$ for the disk. With the
ESO-LV images we confirmed also that the stellar disk of ESO 323-G064
is in the LSB regime. Its  surface brightness
in the B-band ranges from $\sim$23.3 mag arcsec$^{-2}$ at the center
to $\sim$26.5 mag arcsec$^{-2}$ at 35''.  Even though the ESO-LV
images have been useful to get an estimate of the surface brightness,
they had insufficient signal-to-noise and spatial resolution to get
reliable values of the bulge parameters.  We therefore used our VIMOS
observations, collapsing the observed data cube along the dispersion
direction and deriving a surface brightness profile from the resulting
image. We fit a de Vaucouleurs law to this profile and derived an
effective radius $R_e = $ 0\farcs9 (see Figure \ref{fig.photometry})
and a ellipticity ranging from 0.1 to 0.2, not too far from the
adopted spherical approximation.  Although the signal-to-noise
ratio of the VIMOS observations is low, the advantage is that
the VIMOS spatial resolution (0\farcs67 / pixel) is higher than that
of the ESO-LV images (1\farcs35 / pixel). This makes us more confident
of bulge parameter values based on VIMOS observations than values
based on ESO-LV images.

The de Vaucouleurs profile resembles the \citet{Hernquist90}
mass model in which the intrinsic density distribution is given by:

\begin{equation}
\rho_k(r) = \frac{M_L a}{2\pi}\frac{1}{r\left(r+a\right)^3}
\label{eqn.hernquist}
\end{equation}
where $M_L$ is the total luminous mass and $a=R_e/1.8153$.

The projection on the sky of Equation \ref{eqn.hernquist} is given by:

\begin{equation}
\Sigma(R)=2\int_R^{\infty} \frac{\rho_k(r) r}{\sqrt{r^2-R^2}} dr
\label{eqn.surf.proj}
\end{equation}

Equations \ref{eqn.hernquist} and \ref{eqn.surf.proj} will be used in
Equation \ref{eqn.velocity.moment} to determine the velocity moments
for the mass calculation.

\subsection{Intrinsic and projected rotation curve}
\label{sec.rotation.curve}

Because of the limited spatial extent and resolution of the
observations, a ``bidimensional'' approach to the Jeans equation
cannot be reliably carried out. We therefore decided to average
azimuthally and then bin radially the bidimensional velocity and
velocity dispersion fields. To do that, we divided them in $N$ radial
bins\footnote{ Bins have an elliptical shape with $<\epsilon>=0.15$ in
order to take into account the bulge ellipticity we observed.}
(after several tests, it turned out that 1\farcs5 per bin is an
optimal value) and for each radial bin at a given position $r_N$ we:

\begin{itemize} 

\item{} calculated the velocity $V(r_N)$ by fitting the function
$v'(\phi)=V(r_N)\cdot\cos(\phi+P.A.)+Vsys$ to all the velocity data
points within the $N-th$ bin. $\phi$ is the angle measured in the
galaxy equatorial plane, $P.A.$ is the galaxy position angle on the sky
\footnote{We fixed the position angle to the value $P.A. = 38$
    derived in Section \ref{sec:largeRadiiGas}. We tested this
    assumption leaving $P.A.$ free to vary in the fit and we found the
    value to be consistent with 38$^{\circ}$.} and $V_{sys} = 14825$
    \kms\ is the adopted systemic velocity.

\item{} calculated the velocity dispersion $\sigma_N(r_N)$ by
averaging all the velocity dispersion data points within the $N-th$
bin. In this computation, we took into account also the value of the
velocity gradient along the resolution element (1\farcs5) and removed
it from the velocity dispersion.

\end{itemize}

In Equation \ref{eqn.jeans}, the quantity $V_{rot}$ is the intrinsic
stellar rotation curve. It can be parametrized with the expression:

\begin{equation}
V_{rot}(r) = \frac{v_{\infty}r}{\sqrt{r^2+r_h^2}}
\label{eqn.intrinsic.rotation}
\end{equation}
where $v_{\infty}$ is its asymptotic value for the velocity and $r_h$
is a scale parameter. The adopted function was chosen in order to
properly match the observed velocity curve with the minimum number of
free parameters. This expression has been used for example also in the
spherical PNe system of Centaurus A \citep{Hui+95, Peng+04}.

The observed rotation curve is obtained projecting Equation
\ref{eqn.intrinsic.rotation} on to the sky \citep{Binney+87}:

\begin{equation}
V_s(R)=\frac{2R}{\Sigma(R)}\int_R^{\infty} \frac{\rho_k(r) V_{rot}(r)}{\sqrt{r^2-R^2}} dr
\label{eqn.projected.rotation}
\end{equation}

Equations \ref{eqn.intrinsic.rotation} and
\ref{eqn.projected.rotation} will be used in Equations
\ref{eqn.projected.dispersion} and \ref{eqn.velocity.moment} to
compute the velocity moments.

\subsection{Radial mass profile}
\label{sec:radial_mass_profile}

The radial distribution of the total mass $M(r)$ of the galaxy bulge
is given by the sum of the luminous and dark matter contents
\footnote{In Section \ref{sec:dynamical_modeling} we assumed that the
disk contribution to the bulge mass and dynamics is neglegible.}. The
contribution $M_k$ of the luminous component is obtained by
integrating Equation \ref{eqn.hernquist} in the volume. It leads to:

\begin{equation}
M_k(r)=\frac{M_Lr^2}{\left(r+a \right)^2} 
\label{eqn.lum.mass}
\end{equation}

Together with the self consistent case, in which the total mass of the
galaxy is given only by the contribution of the stars, we explored
also two different scenarios for the dark matter content:
\citet[NFW~hereafter]{Navarro+97}, and the pseudo isothermal halos.

Their mass density distribution are:

\begin{equation}
\rho_{NFW}(r)=\frac{\rho_s}{r/r_s \left( r/r_s+1\right)^2}
\label{eqn.nfw.density}
\end{equation}

\begin{equation}
\rho_{isot}(r)=\frac{\rho_0}{1+\left(r/r_0\right)^2}
\label{eqn.isot.density}
\end{equation}
where $r_s$, $\rho_s$, $r_0$ and $\rho_0$ are scale parameters.

The corresponding mass distributions are given by the volume
integration of Equations \ref{eqn.nfw.density} and
\ref{eqn.isot.density}. We obtain:

\begin{equation}
M_{NFW}(r)=4 \pi \rho_s r_s^3 \left[\ln\left(1+r/r_s\right)-\frac{r/r_s}{1+r/r_s} \right]
\label{eqn.nfw.mass}
\end{equation}

\begin{equation}
M_{isot}(r)= 4 \pi \rho_0 r_0^3 \left[ r/r_0 -\arctan\left(r/r_0 \right)\right]
\label{eqn.isot.mass}
\end{equation}

In the dark matter scenarios, the total mass $M(r)$ is given by adding
to $M_k$ one of the contributions expressed in Equations
\ref{eqn.nfw.mass} or \ref{eqn.isot.mass}.

\subsection{Model results}
\label{sec:model_results}

We fit Equations \ref{eqn.projected.dispersion} and
\ref{eqn.projected.rotation} separately to the velocity $V(r_N)$ and
velocity dispersion $\sigma(r_N)$ data points (see Section
\ref{sec.rotation.curve}). 
The actual fit computation was done using the {\tt MPFIT} algorithm,
as done in Section \ref{sec.gaseous.kinematics}.

One could add constraints to the halo parameter by including the disk
\hb\ rotation curve and the disk surface brightness profile into the
fit. However, given the large uncertainties on the disk photometric
profiles, we decided to focus our mass model only to the bulge
regions. For completeness, we present the models in which the disk
data are used in Appendix A.

\begin{figure}
\vbox{
      \psfig{file=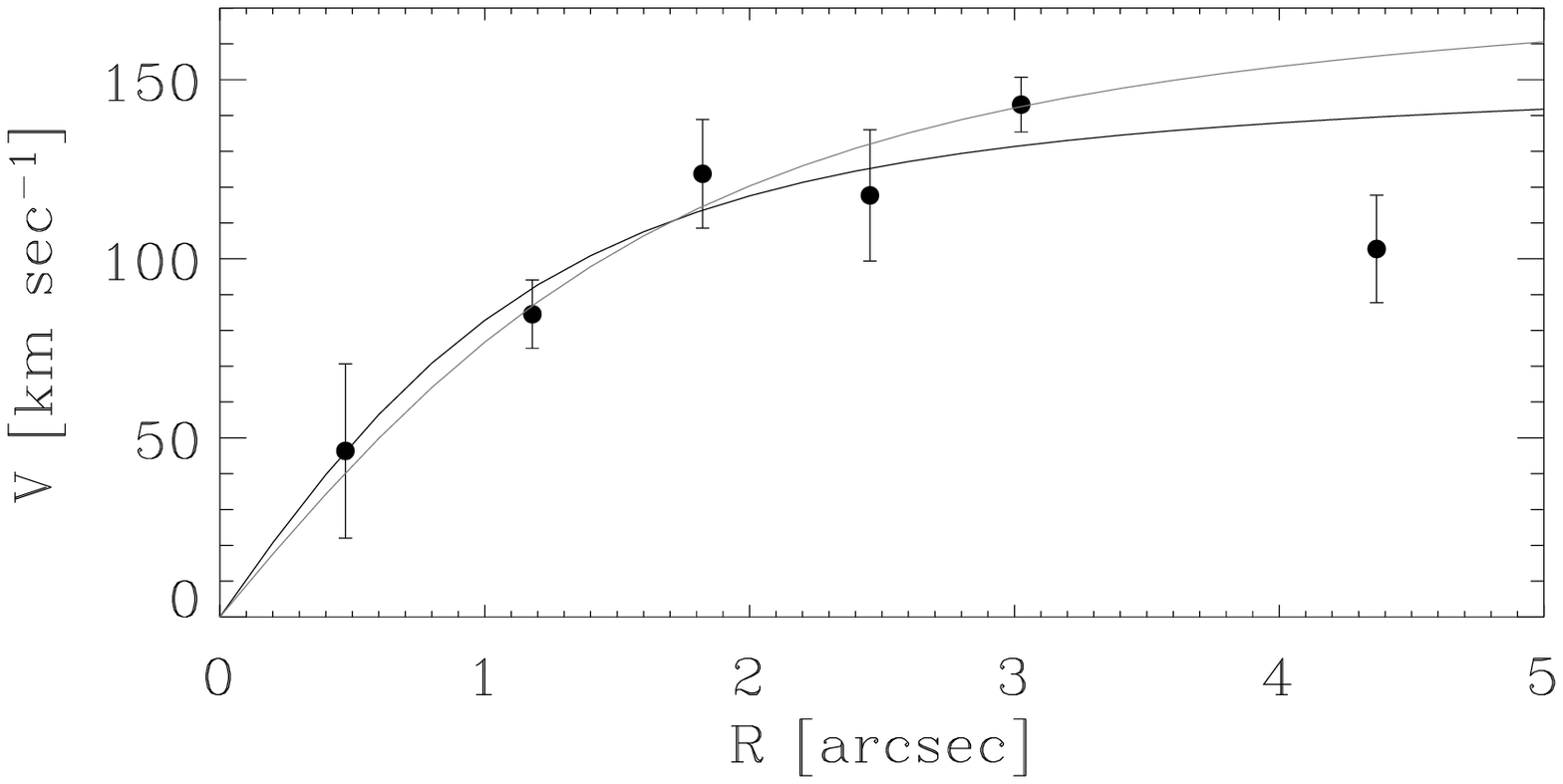,clip=,width=8.0cm}
      \psfig{file=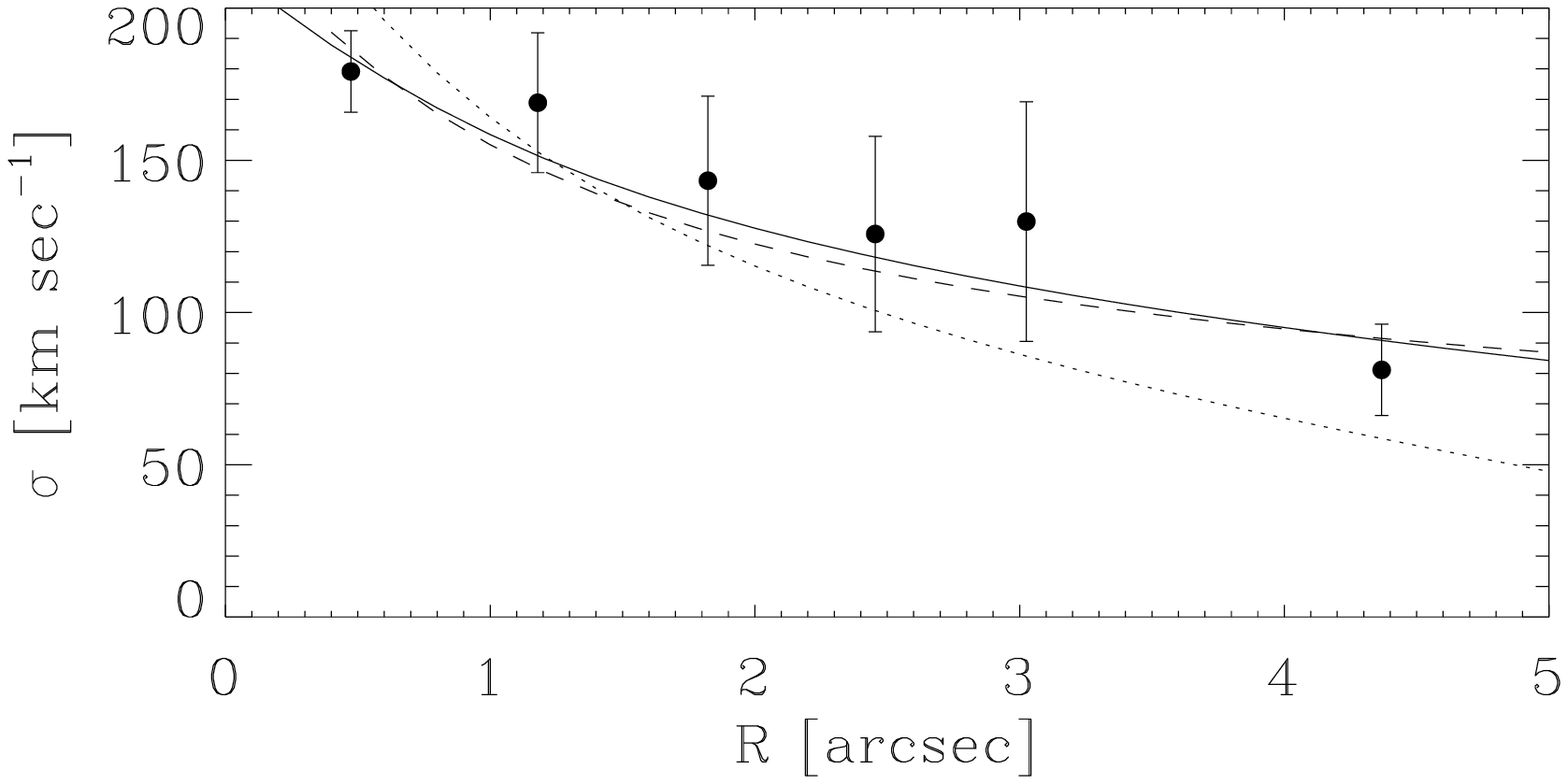,width=8.0cm,clip=}}
\caption{{\it Upper panel:} stellar measured velocity curve of the
bulge ({\it filled circles}, see Section \ref{sec.rotation.curve}) and
the best fit model from Equation \ref{eqn.projected.rotation}.  The
{\it black line} represents the best rotation curve model, while the
{\it gray line} represents the best rotation curve model excluding the
last data point from the fit. {\it Lower panel:} measured radial
velocity dispersion ({\it filled circles}, see Section
\ref{sec.rotation.curve}) compared to the NFW ({\it solid line}), the
pseudo isothermal ({\it dashed line}) and the self consistent ({\it
dotted line}) best fit models.}
\label{fig.model.results}
\end{figure}

The fit results are shown in Figure \ref{fig.model.results} and are
listed in Table \ref{tab.model.result}. 
Figure \ref{fig.model.results} shows the best fit model results
(solid line) together with the observed radial profiles of velocity
and velocity dispersion derived in Section \ref{sec.rotation.curve}.
The last observed point of the rotation curve is below the best fit
rotation curve, possibly the result of the bar influence on the
stellar kinematics for $R\geq$ 3\farcs5 as discussed in Section
\ref{sec:dynamical_modeling}.  Assuming this deviation is entirely due
to the bar, a rough estimate of its effect on our results can be
obtained from a fit of the velocity curve without taking into account
the last velocity measurement. This is shown as the gray line in
the upper panel of Figure \ref{fig.model.results}. The maximum
difference between the two rotation curves (within 4\farcs5) is
$\approx 14\%$. Mass scales as the square of the velocity,
therefore we would expect a maximal underestimation of the total bulge
mass of $\Delta M / M = 2 \Delta V/V = 28\%$, which less than a factor of
1.3.

From Figure \ref{fig.model.results} we can see that the self consistent
model provides a poor fit to the data, compared to the dark matter
scenarios. The bulge kinematics are therefore better explained, under
our assumptions, with the presence of DM, although we are not able to
disentangle between NFW and pseudo isothermal models.

\begin{figure*}
\hbox{
    \psfig{file=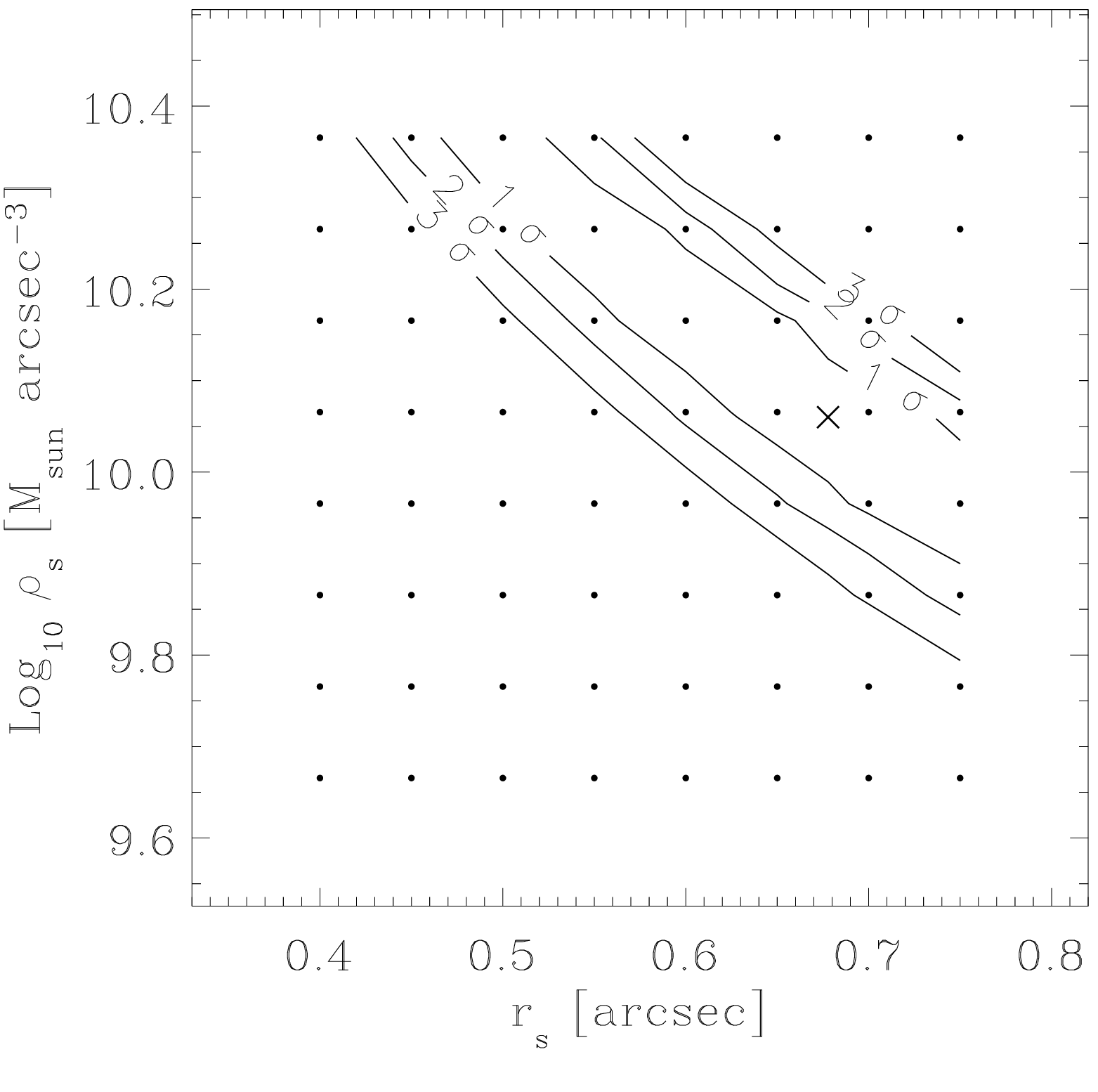,width=5.9cm,clip=}
    \psfig{file=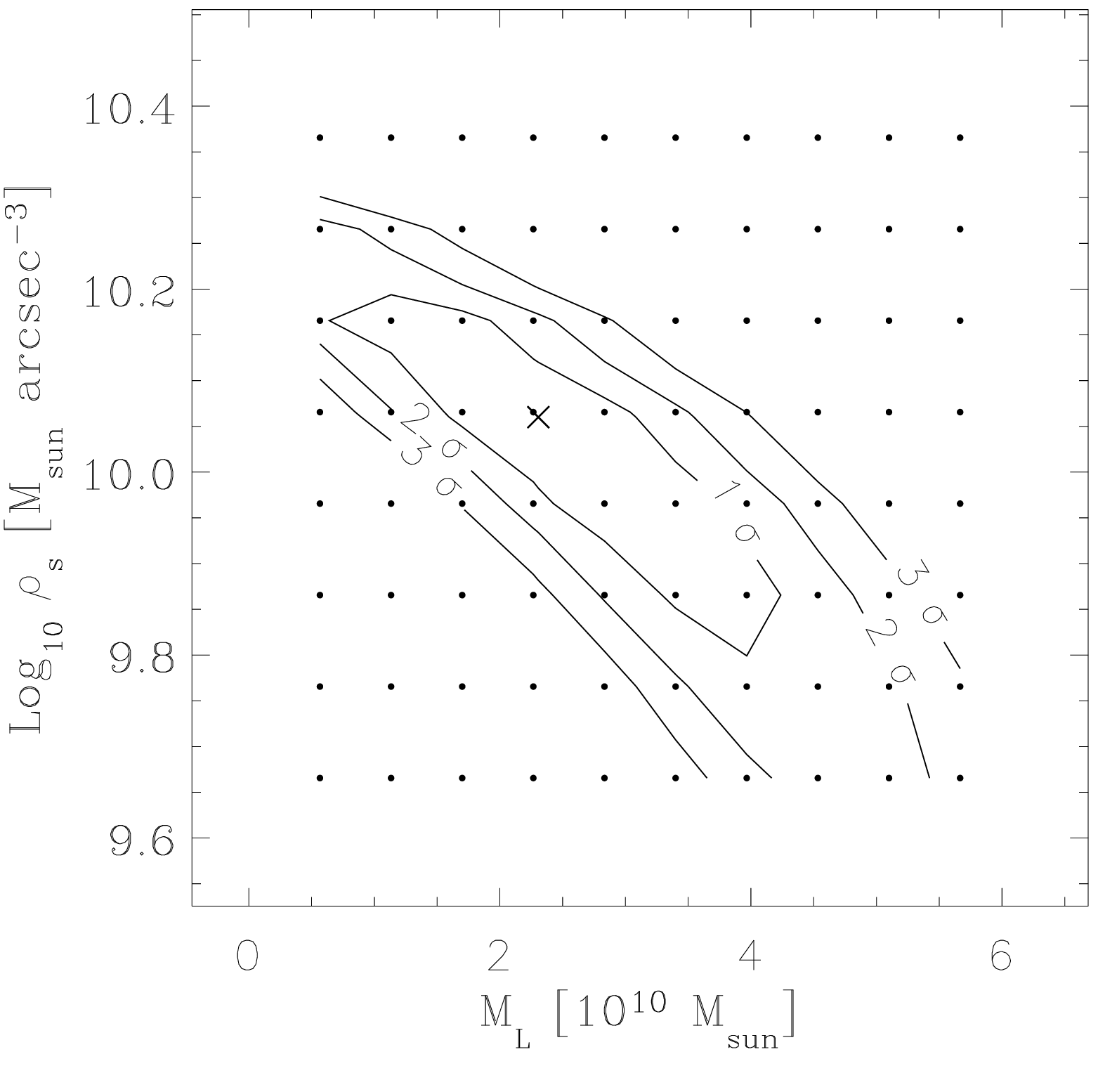,width=5.9cm,clip=}
    \psfig{file=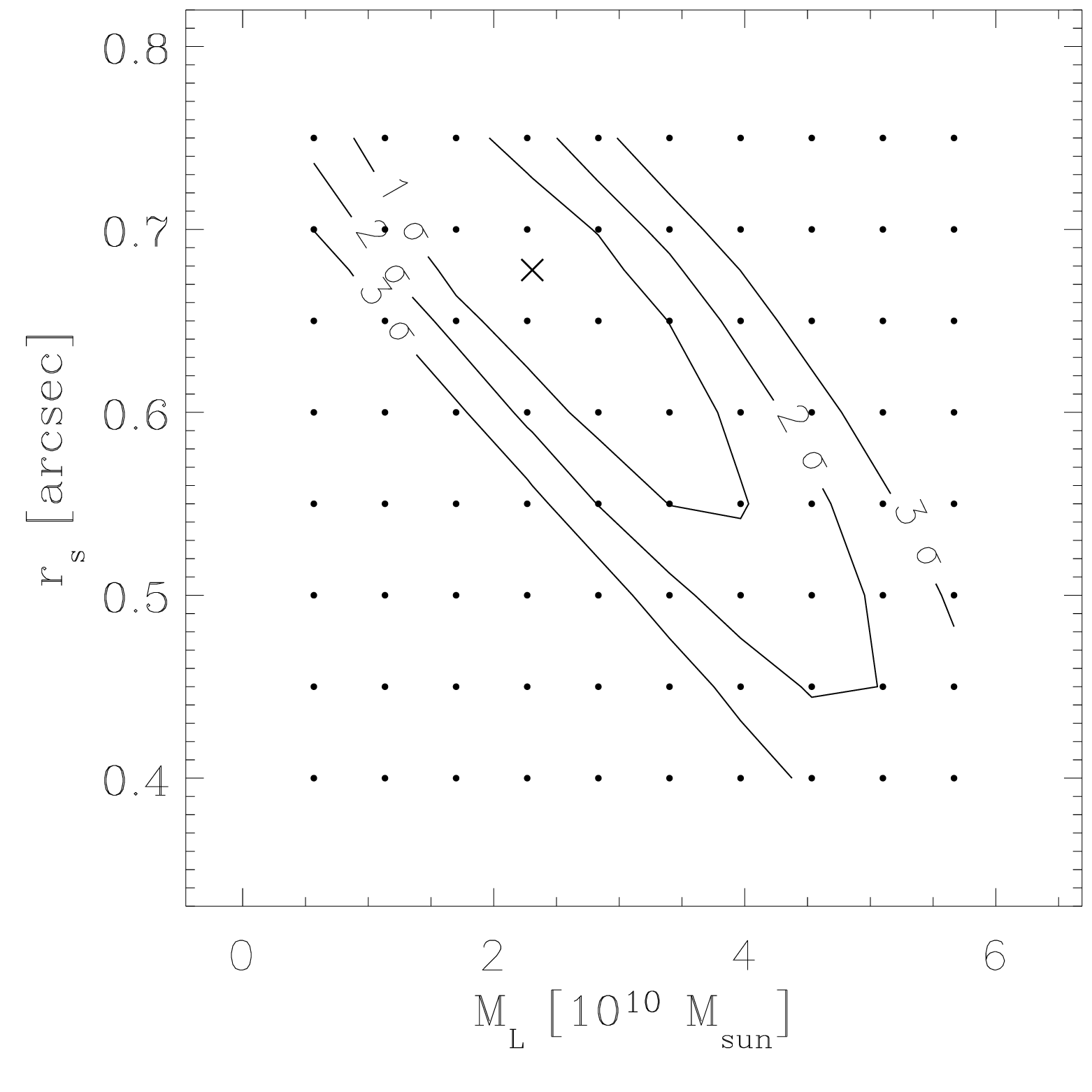,width=5.9cm,clip=} }
\caption{$\chi^2$ confidence levels for the NFW dark matter halo
model. The crosses represent the location of the best fit in the
parameter space ($\tilde{\chi}^2 = 0.8$).}
\label{fig.nfw.grid}
\end{figure*}

\begin{figure*}
\hbox{
    \psfig{file=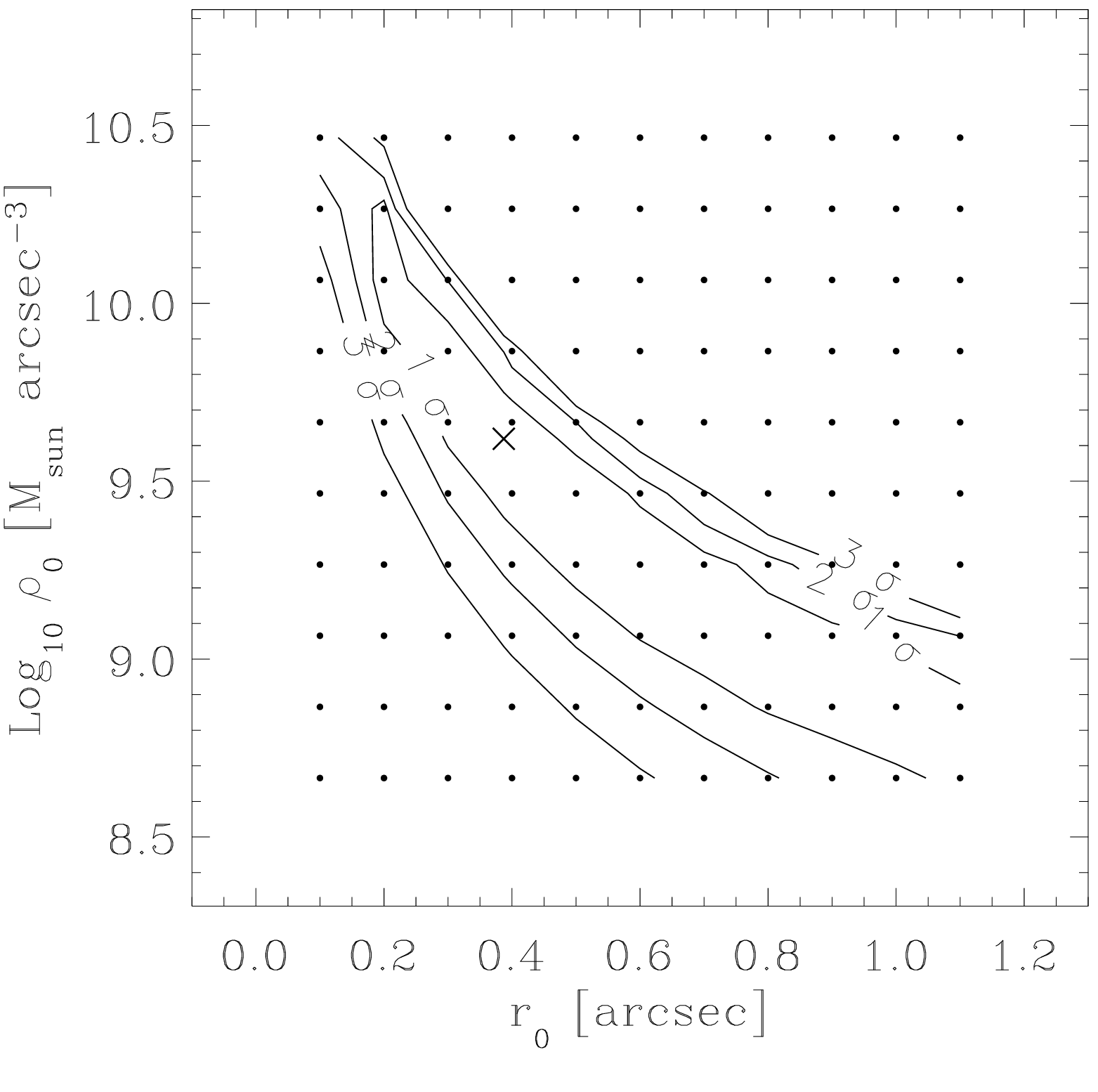,width=5.9cm,clip=}
    \psfig{file=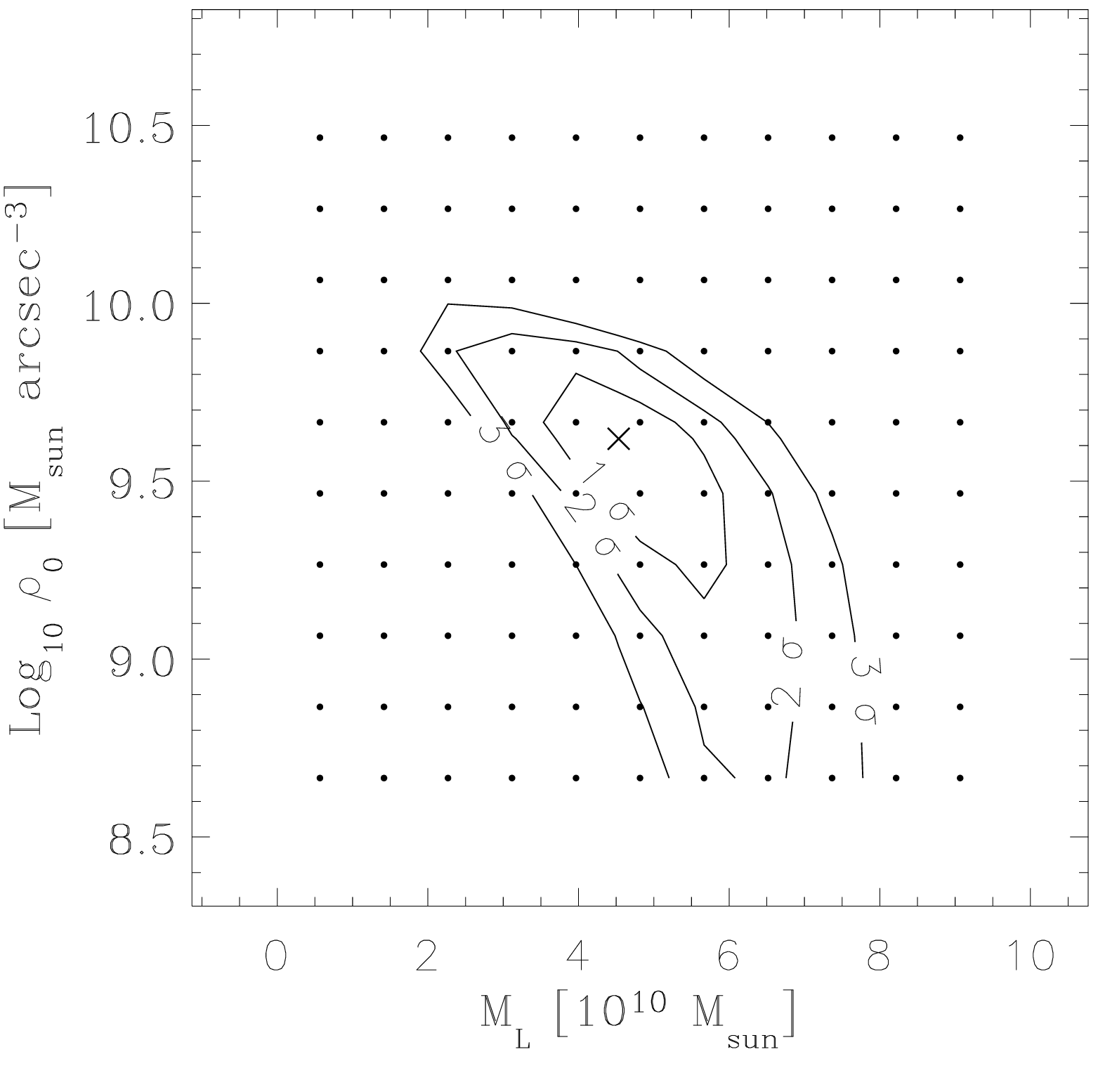,width=5.9cm,clip=}
    \psfig{file=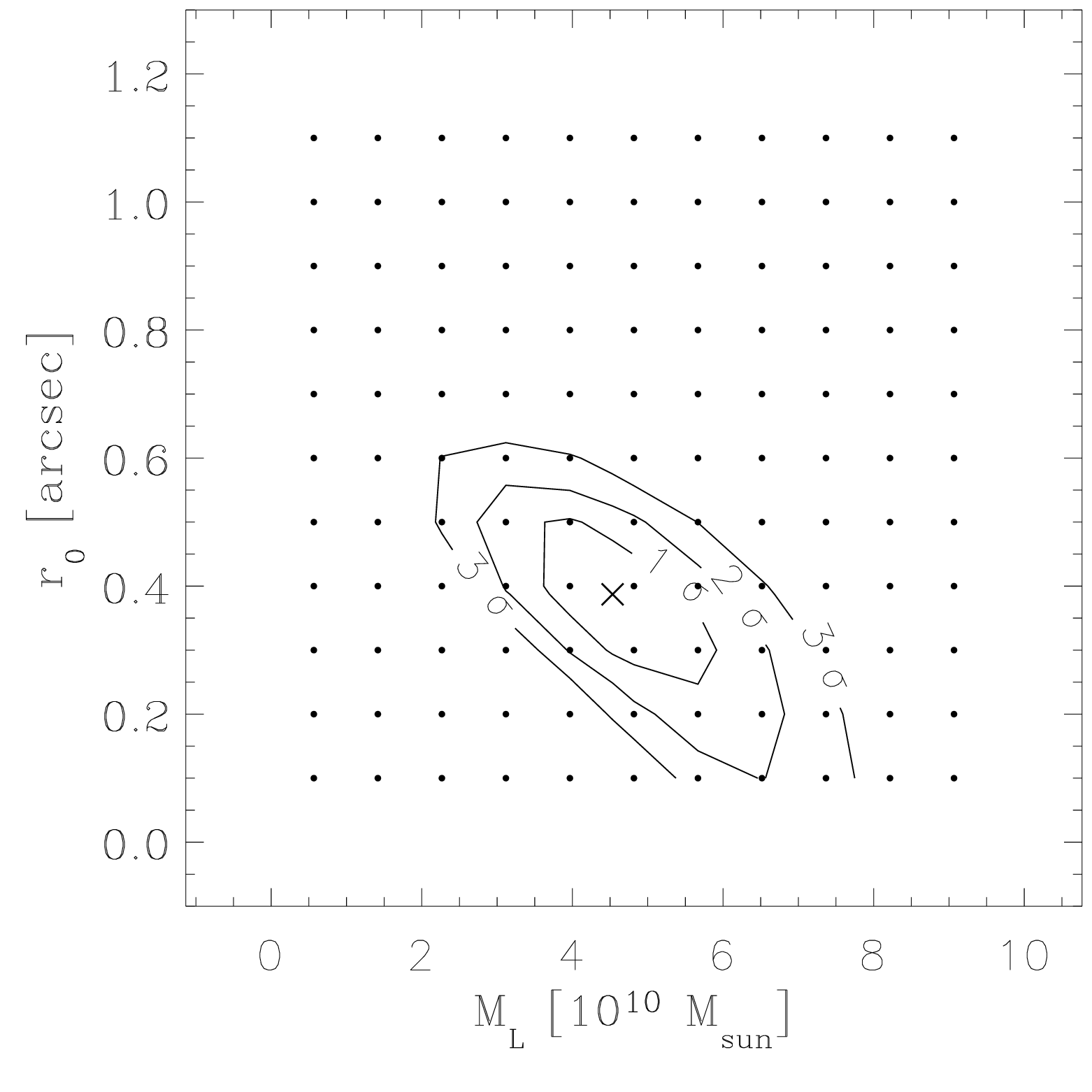,width=5.9cm,clip=}
}
\caption{$\chi^2$ confidence levels for the pseudo isothermal dark
matter halo model. The crosses represent the location of the best fit in
the parameter space ($\tilde{\chi}^2 = 1.5$).}
\label{fig.isot.grid}
\end{figure*}

The reduced $\chi^2$ of both dark matter models is close to 1, while
the self consistent case gives a value of 3.6. In Figures
\ref{fig.nfw.grid} and \ref{fig.isot.grid} we show the $1\sigma$,
$2\sigma$ and $3\sigma$ $\chi^2$ confidence levels in the parameter
space.  Those plots were produced by scaling the measured $\chi^2$
value to the ideal one $\chi^2 = N-M =3$ ($N=6$ is the number of data
points, $M=3$ is the number of free parameters) and taking the
expected $\Delta \chi^2$ variations for 3 parameters (i.e. 3.53, 8.02
and 14.2; \citealt[chap.~15.6]{Press+92}).

\begin{figure}
\psfig{file=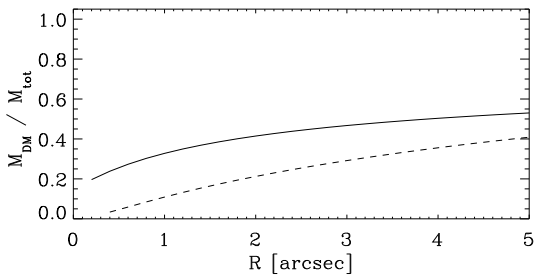,width=8.0cm,clip=}
\caption{Fraction of dark matter mass compared to total mass, as
    a function of radial distance, for the inner $5''$ of ESO
    323-G064. NFW ({\it solid line}); pseudo isothermal ({\it dashed
      line}).}
\label{fig:dm_ratio}
\end{figure}

The total mass of the bulge (calculated using Equations
\ref{eqn.nfw.mass} and \ref{eqn.isot.mass} for $r=5''$) is $(7.4 \pm
3.2) \cdot 10^{10}$ M$_{\odot}$ and $(7.1 \pm 3.6) \cdot 10^{10}$ M$_{\odot}$
according to the NWF and pseudo isothermal scenarios, respectively,
while the ratio of the dark matter content to the total mass is
about 0.55 and 0.42, at that radius.  In the no dark matter scenario
the bulge mass is $(6.5 \pm 1.6) \cdot 10^{10}$ \msun.  The total
fraction of dark matter in the bulge, as a function of the radius, is
 shown in Figure \ref{fig:dm_ratio}.

Errors of the bulge mass are computed by applying error propagation
formulas to Equations \ref{eqn.lum.mass}, \ref{eqn.nfw.mass} and
\ref{eqn.isot.mass}.

\subsubsection{Mass density radial profile}
\label{sec:mass_density}

From the measured stellar rotation curve and the radial velocity
dispersion profiles (i.e. data points in Figure
\ref{fig.model.results}) we derived the mass density profiles and
compared them with the best model predictions.

First we calculated the circular velocity $V_C$ using the asymmetric
drift correction (assuming isotropy):

\begin{equation}
V_C^2(r_N) -  v^2(r_N) = -\sigma^2(r_N)\frac{\partial \ln (\nu)}{\partial \ln (r_N)}
\end{equation}

If we insert the observed bulge $r^{1/4}$ radial profile for the light
distribution $\nu$ of the kinematic tracer we obtain:

\begin{equation}
V_C^2(r_N) =  v^2(r_N) + 1.92 \sigma^2(r_N) \left(\frac{r_N}{R_e}\right)^{1/4}
\end{equation}

Stellar mean circular velocity obtained with the asymmetric drift
correction is $V_c=243 \pm 6$ \kms\, which is consistent with the one
measured from the gas in the disk regions ($248 \pm 6$ \kms).

Finally we calculated the mass density as done in \citet{deBlok+01}
assuming a spherical mass distribution:

\begin{equation}
\rho(r_N)=\frac{1}{4 \pi G} \left[ \frac{V_C^2}{r_N^2} + 2\frac{V_C}{r_N}\left(\frac{\partial V_C}{\partial R} \right)_{R=r_N}\right]
\label{eqn.mass.density.derived}
\end{equation}

\begin{figure}
    \psfig{file=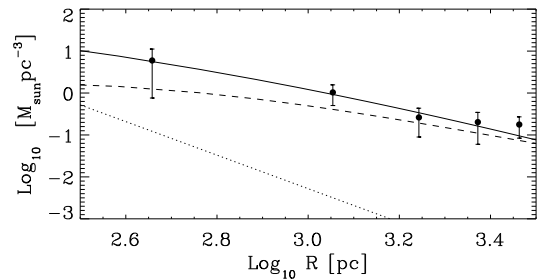,width=8.0cm,clip=}
\caption{{\it Filled circles:} mass density values derived from the
observed velocity and velocity dispersion (Equation
\ref{eqn.mass.density.derived}) compared to the best fit NFW ({\it
continuous line}), pseudo isothermal ({\it dashed line}) and self
consistent ({\it dotted line}) model predictions.}
\label{fig.density.profile}
\end{figure}

We did not fit these {\it calculated} mass density values because they
do not contain more information than the {\it observed} $V$ and
$\sigma$ themselves. Moreover, our direct fit to the observed
quantities does not depend on the assumption adopted in the mass
density calculation.
We show in Figure \ref{fig.density.profile} the comparison between the
derived $\rho$ and the prediction by the NFW and pseudo isothermal
dark matter halo models.

\begin{table}
\centering
\caption{Bulge parameters from the best fit model. Values are
converted assuming $D=194$ Mpc on the right column.}
\begin{tabular}{l c c c}
\noalign{\smallskip}
{\bf Parameter}           &   {\bf Value}                                         &           \\
\noalign{\smallskip}
\hline
\noalign{\smallskip}
{\bf Rotation curve}:     &                                                                                         \\
\noalign{\smallskip}
\ $v_{\infty}$             &   $174 \pm 26$ [\kms]                      &                                     \\
\noalign{\smallskip}
\noalign{\smallskip}
\ $r_h$             &     $1.6 \pm 0.8$ [arcsec]                        &  ($1.5 \pm 0.8$ [kpc])                \\
\noalign{\smallskip}
\hline
\noalign{\smallskip}
\noalign{\smallskip}
{\bf Self consistent}     &  $\tilde{\chi^2} = 3.6$\\
\noalign{\smallskip}
\ $M_L$             & $6.6^{+1.9}_{-1.4}$ [$10^{10}{\rm M}_{\odot}$]    &                                     \\
\noalign{\smallskip}
%\noalign{\smallskip}
%\noalign{\smallskip}
\noalign{\smallskip}
{\bf NFW}:                &    $\tilde{\chi^2} = 0.8  $                       &                                     \\
\noalign{\smallskip}
\ $M_L$             &  $2.3^{3.5}_{-2.5}+$ [$10^{10}{\rm M}_{\odot}$]   &                          \\
\noalign{\smallskip}
\noalign{\smallskip}
\ $\log_{10}\rho_s$ &  $10.1^{+0.2}_{-0.2}$  [$\log_{10}$ (M$_{\odot} {\rm /``}^{3}$)]&  ($\rho_s=15^{+9}_{-5}$ [M$_{\odot}$ pc$^{-3}$]) \\
\noalign{\smallskip}
\noalign{\smallskip}
\ $r_s$             &  $0.7^{+0.2}_{-0.2}$ [arcsec]                     & ($0.67 \pm 0.19$ [kpc])   \\
\noalign{\smallskip}
%\noalign{\smallskip}
%\noalign{\smallskip}
\noalign{\smallskip}
{\bf Isothermal}:         &  $\tilde{\chi^2} = 1.5 $                          &     \\
\noalign{\smallskip}
\ $M_L$             &  $4.5^{+3}_{-2}$ [$10^{10}{\rm M}_{\odot}$]       &                \\
\noalign{\smallskip}
\noalign{\smallskip}
\ $\log_{10}\rho_0$ &  $9.6^{+0.2}_{-0.4}$ [$\log_{10}$ (M$_{\odot} {\rm /``}^{3}$)] & ($\rho_0=5^{+2}_{-1}$  [M$_{\odot}$ pc$^{-3}$])   \\
\noalign{\smallskip}
\noalign{\smallskip}
\ $r_0$             &  $0.4^{+0.1}_{-0.1}$ [arcsec]                    &($0.38 \pm 0.09$ [kpc] )                   \\
\noalign{\smallskip}
\hline
\noalign{\smallskip}
\end{tabular}
\label{tab.model.result}
\end{table}

\section{Discussion and conclusions}

We presented the two-dimensional velocity and velocity dispersion
fields for the gaseous and stellar components of the LSB galaxy ESO
323-G064. The gas emission lines show a very bright and complex
structure within the central $5''$, characterized by 3 peaks, which
we interpret as due to the presence of 3 spatially unresolved
emitting regions. At radii out to $30''$, in the region dominated by the
galaxy disk, a weak \hb\ emission is detected showing a regular
velocity field with a maximal amplitude of $248\pm6$ \kms.

The stellar absorption lines are detectable only in the very innermost
regions of the bulge.  The stellar kinematics shows a regular rotation
with an amplitude of $\pm 140$ \kms\ and a central velocity dispersion
of $\approx 180$ \kms. The intrinsic $(V/\sigma)_{INTR}$ ratio for the
bulge is $0.56 \pm 0.02$, which is consistent with an isotropic
rotator in the $0.1 < \epsilon_{OBS} <0.2$ observed ellipticity range,
considering an inclination of 62 degres. These values of $V/\sigma$
and $\epsilon$ place the bulge of ESO 323-G064 among fast rotator
bulges \citep{Cappellari+06}. The value of $V/\sigma $ is considerably
small when compared to the average value determined from the sample of
6 bulge dominated LSB galaxies of Pizzella et al. 2008 ($<V/\sigma> >
1$).

The circular velocity, $248 \pm 6$ \kms\ measured from the gaseous
disk, places ESO 323-G064 in good agreement with the location of LSB
galaxies in the $V_C -\sigma_c$ plane \citep{Courteau+07b, Courteau+07}.
On the other hand, this value is lower then the value of 320 \kms
predicted by \citet{Pizzella+05}.

The intrinsic bulge ellipticity value for ESO 323-G064 ($\epsilon_{INTR}=0.20
\pm 0.07 $) is consistent with the mean value of bulges of the high surface
brightness disk galaxies $<\epsilon_{HSB}>=0.15$ (as determined by
\citealt{Mendez-Abreu+08}) while it is lower than the average value
for bulge dominated LSB galaxies $<\epsilon_{LSB}>=0.45$ (as found by
\citealt{Pizzella+08} in a small sample of 6 bulge dominated LSB).

The amplitude of the gaseous rotation curve ($248\pm6$ \kms) leads to
an estimate of the total baryonic mass in the galaxy of 
$M_{bar}=(1.9\pm0.2)\cdot 10^{11}$ \msun using the
empirical baryonic Tully Fisher, as done in \citet{McGaugh05}.
Moreover, under the hypothesis of a $\Lambda$CDM universe (see Section
\ref{sec:largeRadiiGas} for details on the assumed parameters), we
estimate the total mass for the dark matter halo $M_{DM} \sim 5 \cdot
10^{12}$ M$_{\odot}$.

We produce spherical isotropic Jeans models for the stellar kinematics
in the bulge, exploring the self consistent, NFW and pseudo isothermal
scenarios. Even though the data are to be taken with some caveats (due
to the lack of good photometry, limited spatial extension and
resolution of the stellar kinematics) with this simple analysis we
show that dark matter scenarios fit the data better than the self
consistent model. The derived total bulge mass is $(7 \pm 3)\cdot
10^{10}$ \msun\ but we are not able to disentangle between the two
different dark matter models.

The derived central bulge mass density (see Table
\ref{tab.model.result}) is $\rho =15^{+9}_{-5}$ [M$_{\odot}$
pc$^{-3}$] in the NFW scenario, and $\rho =5^{+2}_{-1}$ [M$_{\odot}$
pc$^{-3}$] in the pseudo isothermal scenario.
Typical values of central mass density range from few $10^{-3}$
M$_{\odot}$ pc$^{-3}$ to few $10^{-2}$ M$_{\odot}$ pc$^{-3}$ for
regular low surface brightness (see for example \citealt{Kuzio+06,
deBlok+01}) and giant low surface brightness galaxies (\citealt{Pickering+97}).
On the contrary, a much wider range of values is measured for regular
galaxies, from few $10^{-3}$ M$_{\odot}$ pc$^{-3}$ to several $10^3$
M$_{\odot}$ pc$^{-3}$ (i.e. \citealt{Salucci+01,
  Noordermeer+07}). Therefore, in this picture, the bulge of ESO
323-G064 resembles more the central mass density of regular bulges
than those measured in low surface brightness galaxies.
This is consistent also with the fact that bulges
of giant LSB galaxies are photometrically similar to those of
regular high surface brigthtness galaxies \citep{McGaugh+95,
Beijersbergen+99}

\noindent {\bf Acknowledgments.} The authors wish to thank the referee
A. Bosma for useful suggestions which improved the paper content and
the discussion.

\bibliography{coccato2008} 

\section*{Appendix A: Constraining the halo parameters with disk \hb\ velocity curve and surface brightness}
\label{app:A}

The \hb\ velocity curve measured in the disk between $10''<R<30''$
(Section \ref{sec:largeRadiiGas}) and the disk surface photometry from
the ESO-LV images can be used to constrain the dark halo parameters,
under the assumption that the measured \hb\ velocity curve is a good
representation of the galaxy circular velocity.

\subsection*{A1. Circular velocity for the disk and the halo}
The circular velocity predicted for a pseudo isothermal halo is:

\begin{equation}
V_{isot} (r)= V_H \sqrt{1 - \frac{r_0}{r} \arctan \left( \frac{r}{r_0}\right)}
\label{eqn:ISO_V_circ}
\end{equation}
where $V_H=\sqrt(4\pi G \rho_0 r_0^2)$ and $\rho_0$, $r_0$ are the halo
parameters defined in Section \ref{sec:radial_mass_profile}.

The circular velocity predicted for a NFW halo is \citep{Navarro+97}:

\begin{equation}
V_{NFW}(r) = V_{200} \left[ \frac{1}{x} \cdot \frac{\ln(1+cx)-cx/(1+cx)}{\ln(1+c)+c/(1+c)}   \right]^{1/2}
\label{eqn:NFW_V_circ}
\end{equation}

In Equation \ref{eqn:NFW_V_circ} $c$ is the concentration parameter,
related to the density parameter $\rho_s$ (used in our fit procedure)
and the critical density $\rho_{critic} = 3H^2/(8\pi G)$ ($H=75$ is
the adopted value for the Hubble constant) with the following
Equation:

\begin{equation}
\rho_s / \rho_{critic} = \frac{200}{3}\frac{c}{\ln(1+c)-c/(1+c)}
\end{equation}

The parameter $V_{200}$ in Equation \ref{eqn:NFW_V_circ} is the
circular velocity at $r_{200}=r_s c$ ($r_s$ is the halo scale
parameter used in our fit procedure) and it is defined as: 

\begin{equation}
V_{200}=\left( \frac{G M_{200}}{r_{200}} \right)^{1/2}
\end{equation}
where
\begin{equation}
M_{200}=200 \frac{4\pi}{3} \cdot  \rho_{critic} r_{200}^3.
\end{equation}

Together with the halo circular velocities defined by Equations
\ref{eqn:ISO_V_circ} and \ref{eqn:NFW_V_circ} we have to add the
contribution given by the stellar disk (the bulge contribution is
negligible at this radial range). Assuming an exponential disk with
central mass surface density $M_0$ and scale length $h$, the
predicted circular velocity is \citep{Freeman70}:

\begin{equation}
V_{disk} (r)= \sqrt{ 4 \pi G M_0 \frac{r^2}{4h}\left[ I_0(\frac{x}{2h})K_0(\frac{x}{2h}) - I_1(\frac{x}{2h})K_1(\frac{x}{2h})\right] } 
\label{eqn:DISK_V_circ}
\end{equation}
where $I_n$ and $K_n$ are modified Bessel functions of the first and second kind.

To obtain the mass surface density $M_0$ and the disk scale parameter
we retrieved ESO-LV images of ESO 323-G064 in the $B$ and $R$ bands
and performed a photometric decomposition with {\tt galfit} \citep{Peng+02}.

The result of the {\tt galfit} decomposition are listed in Table
\ref{tab:galfit}. From the total disk magnitude in $R$ and $B$ band we
compute a color of $(B-R)_{disk}=1.70 \pm 0.5$. Using the prescription
by \citet{Bell+01} we can compute the mass-to-light ratio of the disk
in the $R$ band $(M/L)_{disk} = 4.81 \pm 0.77$. The computation made
use of the relation between color and mass-to-light ratio for
different models as listed in Table 3 of
\citet{Bell+01}. We use the relation in the $R$ because the scatter
between different models is smaller.

\begin{table}
\caption{Galfit best fit parameters.}
\begin{tabular}{l c c }
\hline
  &     $R$-band              &  $B$-band\\
\hline
Bulge:   &                  &        \\
  $M_{tot}$ & 16.23 $\pm$ 0.5 mag &  17.75 $\pm$ 0.5 mag       \\
  $R_e$   & 0.65 $\pm$ 1 arcsec &   0.55 $\pm$  arcsec      \\
  $b/a$  & 0.71 $\pm$ 0.3      &  0.71 $\pm$ 0.3      \\
  $PA$   & 13 $\pm$ 20 deg     &  -72 $\pm$ 20 deg      \\
\noalign{\smallskip}
\hline
 Disk: &                  &            \\
  $M_{tot}$ & 14.11 $\pm$ 0.5 mag & 15.81 $\pm$ 0.5 mag        \\ 
  $\mu_0$ & 21.2 $\pm$ 0.4 mag & 23.3 $\pm$ 0.4 mag        \\ 
  $h$   & 12.4 $\pm$ 1 arcsec & 13.6 $\pm$ 2 arcsec \\
  $b/a^{(*)}$  & 0.53 $\pm$ 0.3      & 0.52 $\pm$ 0.3 \\
  $PA$   & 33 $\pm$ 5 deg      &  32 $\pm$ 6 deg\\
\noalign{\smallskip}
\hline
\noalign{\smallskip}
\end{tabular}\\
\label{tab:galfit}
\begin{minipage}{7cm}
Note: $^{(*)}$ A disk axial ratio $b/a = 0.53$ corresponds to an
inclination $i=60^{\circ}$ , assuming an intrinsic disk axial ratio of
$q_0$=0.18 \citep{Guthrie92}.
\end{minipage} 
\end{table}

From the central disk surface brightness $\mu_0=21.2 \pm 0.3$ mag
arcsec$^{-2}$ measured from the {\tt galfit} decomposition in the
$R$-band, assuming a distance of $D=194$ Mpc, an extinction of
$A_R=0.231$ mag (from the NED database) the disk mass-to-light ratio
$(M/L)_{disk} = 4.81 \pm 0.77$ we calculate the central mass surface
density $M_0=(491\pm201)$ M$_{\odot}$ pc$^{-2}$ = $(4.3 \pm 1.8) \cdot 10^8$
M$_{\odot}$ arcsec$^{-2}$.

Therefore, we can compute the total circular velocity $V_{C}$ by
adding the contribution of the light (Equation \ref{eqn:DISK_V_circ})
to the contribution of the halo (Equations \ref{eqn:ISO_V_circ} or
\ref{eqn:NFW_V_circ} depending on the adopted scenario).

\subsection*{A2. Mass model using stellar and gaseous kinematics}
We have also performed the mass model fit as described in Section
\ref{sec:dynamical_modeling} using the constraints from the \hb\
circular velocity. As in the previous case we fit first the empirical
bulge stellar rotation curve (to obtain $r_\infty$ and $r_h$, Equation
\ref{eqn.intrinsic.rotation}) and then we fit {\it simultaneously} the
stellar velocity dispersion (i.e the filled circles in Figure
\ref{fig.model.results}) and the  \hb\ circular velocity (i.e the open
diamonds in Figure \ref{fig:faint_field}) data.  

\begin{figure}
\vbox{
      \psfig{file=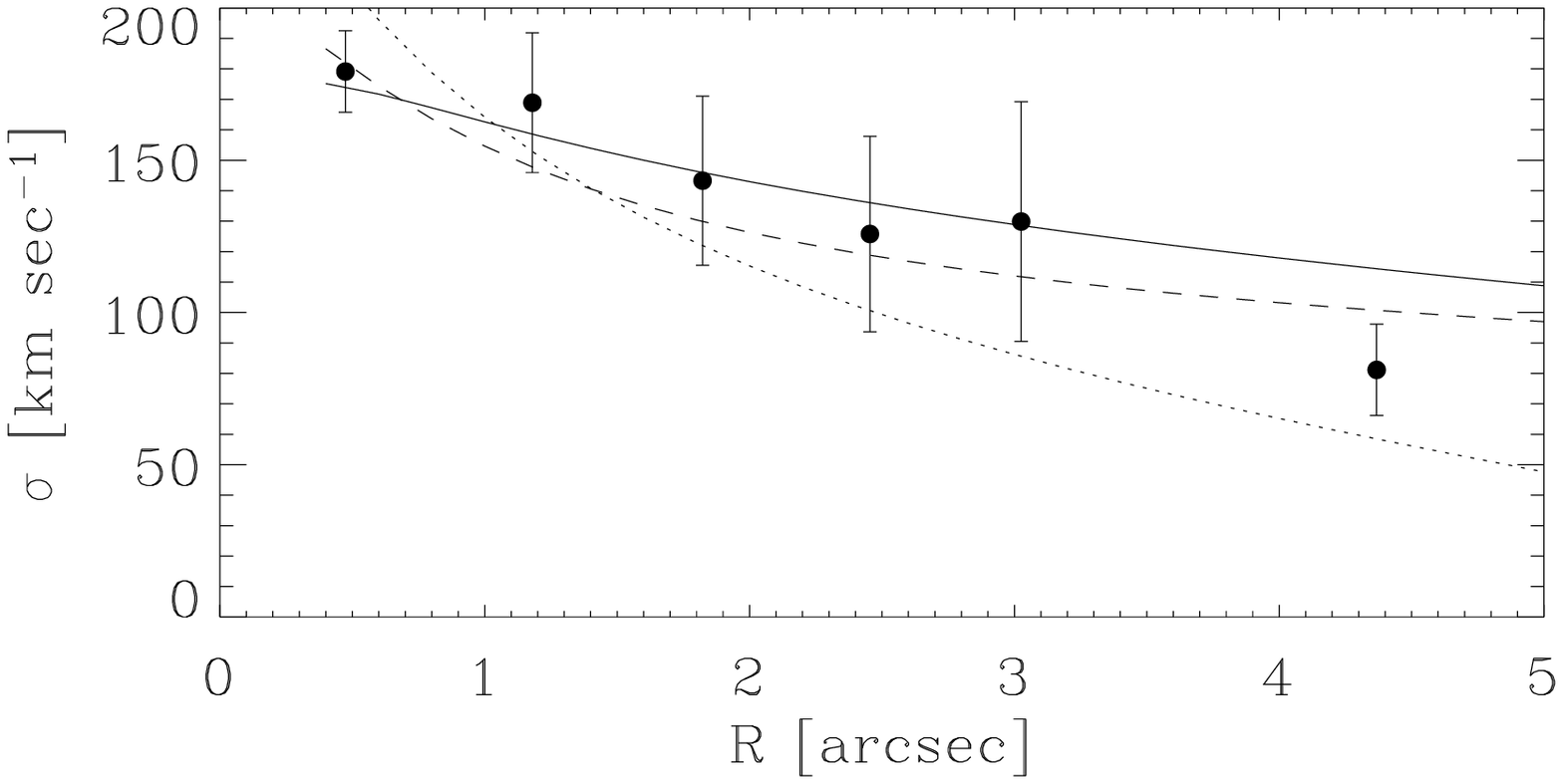,clip=,width=8.0cm}
      \psfig{file=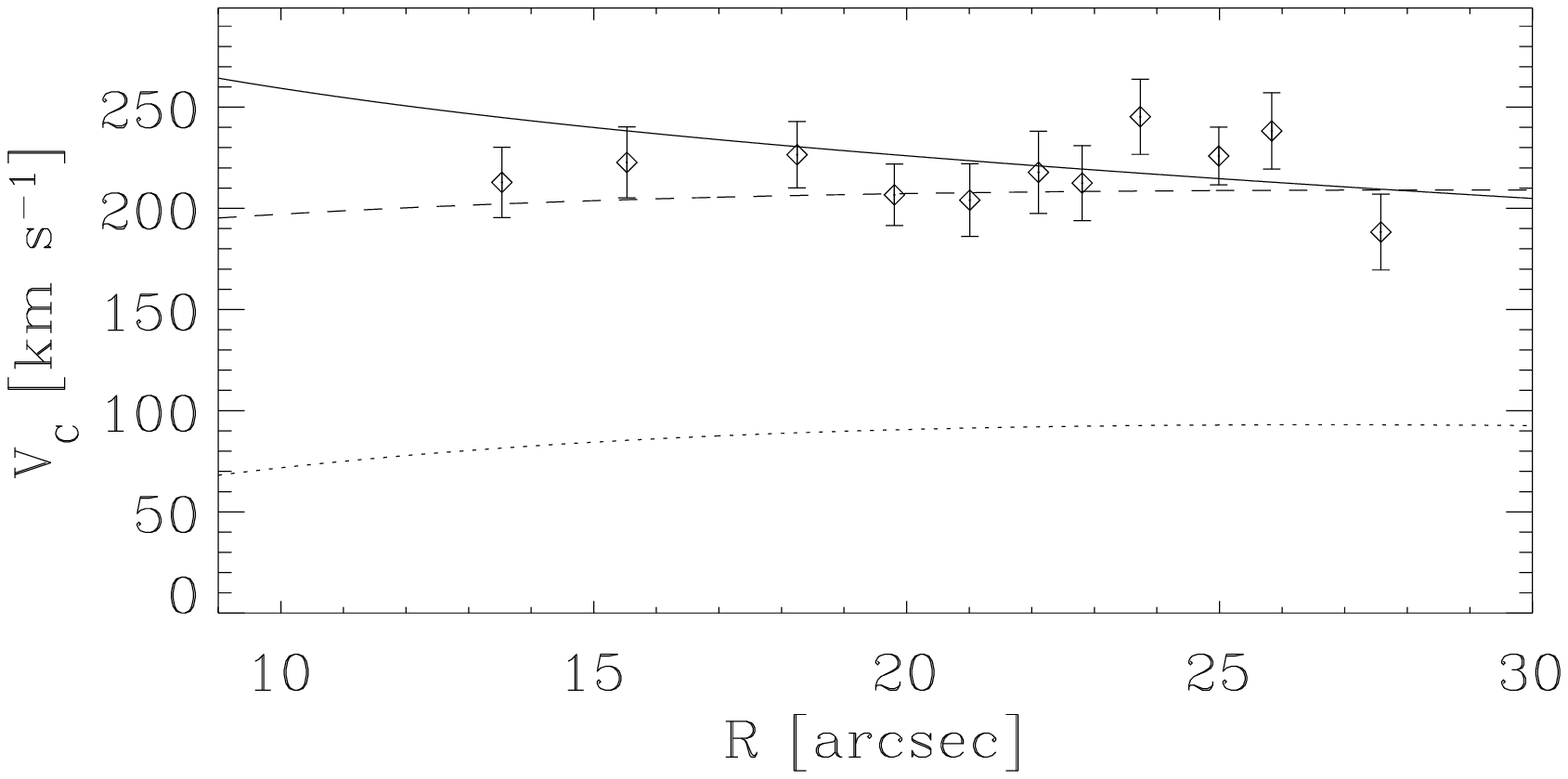,width=8.0cm,clip=}}
\caption{Model results obtained including the disk data. {\it Upper
panel:} stellar velocity dispersion ({\it filled circles}, see Section
\ref{sec.rotation.curve}) compared to the NFW ({\it solid line}), the
pseudo isothermal ({\it dashed line}) and the self consistent ({\it
dotted line}) best fit models. {\it Lower Panel:} measured circular
velocity ({\it open circles}, from the \hb\ rotation curve derived in
Section \ref{sec:largeRadiiGas}) compared to the NFW ({\it solid
line}), the pseudo isothermal ({\it dashed line}) and the self
consistent ({\it dotted line}) best fit models.}
\label{fig.App.model.results}
\end{figure}

Fit results are shown in Figure \ref{fig.App.model.results} and the
best fit parameters are listed in Table \ref{tab:app_mod_res}. The results
are similar to those in Table \ref{tab.model.result}.
With the additional constraints from the disk, both pseudo isothermal and NFW
scenario give a reasonable fit to the data and it is not possible, given the
errors, to say which is the best one.

\begin{table}
\centering
\caption{Best fit parameters from the simultaneous fit of the stellar velocity
dispersion and the \hb\ rotation curve. Values are converted assuming
$D=194$ Mpc on the right column.}
\begin{tabular}{l c c c}
\noalign{\smallskip}
{\bf Parameter}           &   {\bf Value}                                         &           \\
\noalign{\smallskip}
\hline
\noalign{\smallskip}
{\bf Rotation curve}:     &                                                                                         \\
\noalign{\smallskip}
\ $v_{\infty}$             &   $174 \pm 26$ [\kms]                      &                                     \\
\noalign{\smallskip}
\noalign{\smallskip}
\ $r_h$             &     $1.6 \pm 0.8$ [arcsec]                        &  ($1.5 \pm 0.8$ [kpc])                \\
\noalign{\smallskip}
\hline
\noalign{\smallskip}
\noalign{\smallskip}
{\bf Self consistent}     &  $\tilde{\chi^2} = 3.6$\\
\noalign{\smallskip}
\ $M_L$             & $6.6^{+1.9}_{-1.4}$ [$10^{10}{\rm M}_{\odot}$]    &                                     \\
\noalign{\smallskip}
%\noalign{\smallskip}
%\noalign{\smallskip}
\noalign{\smallskip}
{\bf NFW}:                &    $\tilde{\chi^2} = 0.9  $                       &                                     \\
\noalign{\smallskip}
\ $M_L$             &  $0.54^{1.0}_{-0.4}+$ [$10^{10}{\rm M}_{\odot}$]   &                          \\
\noalign{\smallskip}
\noalign{\smallskip}
\ $\log_{10}\rho_s$ &  $10.0^{+0.2}_{-0.2}$  [$\log_{10}$ (M$_{\odot} {\rm /``}^{3}$)]&  ($\rho_s=13^{+8}_{-5}$ [M$_{\odot}$ pc$^{-3}$]) \\
\noalign{\smallskip}
\noalign{\smallskip}
\ $r_s$             &  $0.85^{+0.14}_{-0.15}$ [arcsec]                     & ($0.80 \pm 0.18$ [kpc])   \\
\noalign{\smallskip}
%\noalign{\smallskip}
%\noalign{\smallskip}
\noalign{\smallskip}
{\bf Isothermal}:         &  $\tilde{\chi^2} = 1.4 $                          &     \\
\noalign{\smallskip}
\ $M_L$             &  $3.5^{+3}_{-2}$ [$10^{10}{\rm M}_{\odot}$]       &                \\
\noalign{\smallskip}
\noalign{\smallskip}
\ $\log_{10}\rho_0$ &  $9.6^{+0.2}_{-0.4}$ [$\log_{10}$ (M$_{\odot} {\rm /``}^{3}$)] & ($\rho_0=5^{+2}_{-1}$  [M$_{\odot}$ pc$^{-3}$])   \\
\noalign{\smallskip}
\noalign{\smallskip}
\ $r_0$             &  $0.39^{+0.08}_{-0.09}$ [arcsec]                    &($0.37 \pm 0.08$ [kpc] )                   \\
\noalign{\smallskip}
\hline
\noalign{\smallskip}
\end{tabular}
\label{tab:app_mod_res}
\end{table}

The formal errors given by the fit algorithm are slightly lower than
the ones listed in Table \ref{tab.model.result}, because of the addition of
data points in the fit. However, the fit errors do not include
the uncertainties from the disk photometry and the errors in the
relation between mass-to-light ratio and disk $(B-R)$ color.

Errors in disk central surface brightness and in the mass-to-light
ratio translate into an error of $\Delta M_0 / M_0 \sim 41\%$ on the
mass central surface density, which is an error of $\sim 20\%$ on the
disk velocity curve ($V_{disk} \propto \sqrt{M_0}$).  However, the
fact that the two different fit approaches lead to similar results is
reassuring.

Deep photometric observations and an accurate determination of the
mass-to-light ratio are highly desirable to better constrain the mass
distribution in ESO 323-G064.

\end{document}